\documentstyle[12pt]{article}
\def\sec{\setcounter{equation}{0}}

\newcounter{popnr}
\addtocounter{popnr}{1}

\newcommand{\beq}{\begin{eqnarray}}
\newcommand{\eeq}{\end{eqnarray}}
\newcommand{\beqq}{\begin{eqnarray*}}
\newcommand{\eeqq}{\end{eqnarray*}}

\newcommand{\bbR}{I\!\!R}
\def\RR{\bbR}
\newcommand{\bbC}{I\!\!\!C}
\newcommand{\bbG}{I\!\!\!G}
\def\GG{\bbG}
\newcommand{\CC}{I\!\!\!\!C}
\def\EE{I\!\!\!E}
\newcommand{\gh}{h\!\!\!\smash h} % ma byc gotyckie h
\newcommand{\dtilde}[1]{\mathop{#1}\limits_{\sim}{}}
\newtheorem{prop}{Proposition}[section]
\newtheorem{lem}[prop]{Lemma}
\newtheorem{df}[prop]{Definition}
\newtheorem{th}[prop]{Theorem}
\newtheorem{cor}[prop]{Corrollary}
\def\i{\item}
\def\bd{\begin{description}}
\def\ed{\end{description}}
\def\b{\beta}
\def\a{\alpha}
\def\d{\delta}
\def\g{\gamma}
\def\G{\Gamma}
\def\^{\widehat}
\def\~{\widetilde}
\def\bel{\begin{equation}\label}
\def\ee{\end{equation}}
\def\r#1{(\ref{#1})}
\def\E{\,^E}
\def\K{{\cal K}}
\def\Sbo{S^\beta_0}
\def\Sbe{S^\beta_\epsilon}
\def\void{\,\cdot\,}
\def\Rd{\bbR^d}
\def\Cbo{C^\beta_0}
\def\Cbe{C^\beta_\e}
\def\ClL{C^\lam_\Lam}
\def\DRd{{\cal D}'(\Rd)}
\def\Drd{{\cal D}(\Rd)}
\def\o{\overline}
\def\tens{\otimes}
\def\n#1{${#1}_1$, \ldots, ${#1}_n$}
\def\proof{{\bf Proof:} \par}
\def\endproof{\\\rightline{$\Box$}\vskip 3mm}
\def\dsp{\displaystyle}
\def\Ep{{\cal E}(p)}
\def\({\left(}
\def\){\right)}
\def\SRd{S(\Rd)}
\def\half{{1 \over 2}}
\def\ba{\begin{array}}
\def\ea{\end{array}}
\def\dh{{d+\half}}
\def\sumno{\sum_{n \geq 0}}
\def\ZZ{Z\!\!\!Z}
\def\Om{\Omega}
\def\remarks{\par {\bf Remarks} \\}
\def\ben{\begin{enumerate}}
\def\een{\end{enumerate}}
\def\Gbo{\Gamma^\beta_0}
\def\intR{\int_0^\infty}
\def\nSb{^n S_\beta}
\def\<{\left<}
\def\>{\right>}
\def\H{{\cal H}}
\def\W{{\cal W}}
\def\Wh{{\cal W}(\gh)}
\def\A{\,^A}
\def\m{\mbox{\bf m}}
\def\ds{\dtilde s}
\def\dW{\dtilde W}
\def\dt{\dtilde \tau}
\def\Ho{{\cal H}_0}
\def\Omo{\Omega_0}
\def\aot#1{\a^0_{t_{#1}}}
\def\mo{\m_0}
\def\dWs#1{(\dW^{#1},\,\ds^{#1})}
\def\Epo{\E\pi_0}
\def\to{\longrightarrow}
\def\p{\partial}
\def\apW#1#2#3#4{\E\a^0_{i#1}(\Epo(W(#2)))\,\ldots\E\a^0_{i#3}(\Epo(W(#4)))}

\def\aPWbi#1#2#3#4#5{#5\a^0_{#1}(#5\pi_0(W(#2)))\,\ldots#5\a^0_{#3}(#5\pi_0(W(#4)))}
\def\apw#1#2#3{#3\a^0_{i#1}(#3\pi_0(W(#2)))}
\def\apwbi#1#2#3{#3\a^0_{#1}(#3\pi_0(W(#2)))}
\def\1{^{-1}}
\def\2{^{\prime\prime}}
\def\3{^{\prime\prime\prime}}
\def\Ah{{\cal A}(\gh)}
\def\:{\,:\,\,\,}
\def\fn#1{\mathop{\rm #1}}
\def\x{\times}
\def\ooo{,\,\ldots,}
\def\LRd{L^2(\Rd)}
\def\D{\Delta}
\def\|{|\!|}
\def\ve{\varepsilon}

\def\xf#1#2{\<\xi_{#1},\,f_{#2}\>}
\def\xn{^{\x n}}
\def\fgf#1#2{\<\phi,#2 g_{#1}\tens f_{#1}\> }
\def\fdf#1{\<\phi, \d_{\tau_{#1}}\tens f_{#1}\> }
\def\e{\epsilon}
\def\apwx#1#2#3#4{\a^\xi_{i#1}(\pi^\xi(W(#2)))\,\ldots\a^\xi_{i#3}(\pi^\xi(W(#4)))}
\def\fff{{f_1\ooo f_n}}
\def\ttt{{\tau_1\ooo \tau_n}}
\def\xot{\xi^0_t}
\def\xt{\xi_t}
\def\xito{(\xot)_{t\in\RR}}

\def\t{\tau}
\def\mbo{\mu^\b_0}
\def\vf{\varphi}
\def\S{\Sigma}
\def\Kb{K_\b}
\def\supp{\mathop{\rm supp}\limits}
\def\inf{\mathop{\rm inf}\limits}
\def\s{\sigma}
\def\Lam{\Lambda}
\def\lam{\lambda}
\def\B{{\cal B}}
\def\om{\omega}
\def\Lb{{(\Lam,b)}}
\def\Lbn{{(\Lam_n,b_n)}}
\def\Ln{\Lam_n}
\def\K{{\cal K}}
\def\limni{\lim_{n\rightarrow\infty}}
\def\dsq{\dsp \quad}
\def\dsqq{\dsp \qquad}
\def\lbl{{(\Lam,b_{\p\Lam})}}

\def\LOm{L^2(\Om)}
\def\Tr{\fn{Tr}}
\def\glam{{\G_{-1}(L^2(\Lam))}}
\def\ra{\rightarrow}
\def\bm{{(\b,\mu)}}
\def\intl{\int\limits}

\def\[{\left[}
\def\]{\right]}
\def\gtf{g \tens f}
\def\ftg{f \tens g}
\def\mlL{\mu^\lam_\Lam}
\def\Im{\fn{Im}}
\def\Re{\fn{Re}}
\def\Vbo{\nu^\b_0}

\begin{document}
 \setcounter{page}{0}
\begin{titlepage}
%\vbox{
\title{Gentle Perturbations of the Free Bose Gas I.
% The Noncritical Regime
}
\author{\em Roman Gielerak\thanks{supported by KBN  Grant No 200609101
 and the EC Mobility Grant 722} and Robert Olkiewicz\\
\\
Institute of Theoretical Physics\\
University of Wroc{\l}aw\\
pl. Maksa Borna 9, 50-204 Wroc{\l}aw, POLAND}
\date{}
\maketitle
\thispagestyle{empty}
\begin{abstract}
It is demonstrated that the thermal structure of the noncritical free
Bose Gas is completely described  by
certain periodic generalized Gaussian stochastic process or equivalently
by certain periodic generalized Gaussian random field. Elementary
properties of this Gaussian stochastic thermal structure have been
established. Gentle perturbations of several types of the free thermal
stochastic structure are studied. In particular new models of
non-Gaussian thermal structures have been constructed and a new
functional integral representation of the corresponding euclidean-time
Green functions have been obtained rigorously.
%\vskip 3mm
\begin{description}
\item[Key words:] free Bose Gas, $W^*$-KMS structure, periodic
generalized stochastic process, gentle perturbations, multitime Green
functions.
\end{description}
\end{abstract}
%}
\end{titlepage}

\newpage
\sec
\section{Introduction}
A variety  of existence and analycity results --- also
constructive ones --- have been rigorously obtained for some realistic
models of nonrelativistic quantum matter in  thermal equilibrium
(see \cite{1,2,3,4,5,6,7,8}). Nevertheless, a number of basic questions on the
origin of the fundamental macroscopic quantum phenomena such as
superconductivity, superfluidity, etc. \cite{9,10} are still out of
rigorous demonstrations in the above mentioned realistic treatment.
Only for mean-field-like and exactly solvable models a
mathematically well defined analysis of these phenomena has been performed
\cite{11,12,13}. It is worthwhile to mention here the recent activity on the
superconductivity problem in Fermi matter models of physical interest
\cite{14,15}, which is based on the rigorous renormalization group
approach invented by Gallavotti and his coworkers \cite{16}.

The main objective of the present series of papers is to construct a class of
models of selfinteracting
nonrelativistic Bose matter in a thermal equilibrium for which a
rigorous discussion of the Bose-Einstein condensation, as well as other
phase transitions, would be feasible. In order to approach this goal, we
intend to use extensively methods inferred from the Constructive
Euclidean QFT. In the first paper of the planned series the stochastic
content of the fundamental $W^*-KMS$ structure of a free, noncritical
Bose gas \cite{17} is described. We prove that the abelian sector
of the Weyl algebra may be
described by a certain generalized periodic stochastic process with
values in $\DRd$ (the space of the Schwartz distributions) and, what is
more, that a reconstruction of the whole thermal structure can be
derived from it (see prop \ref{p2.5} below). Similar situation do also
occur in the case of the critical Bose gas when the underlying process
is nonergodic \cite{18}.
Having described a free Bose gas in terms of stochastic processes, one
may perturb them with multiplicative (-like) functionals, thereby
creating some new non-Gaussian thermal processes. Furthermore, given
such a process, one is able to reproduce its $W^*$ KMS counterpart by
methods of \cite{17,19,20}. In this article we shall confine ourselves to the
simplest case of perturbations, which we have called (after \cite{21})
gentle perturbations of a free thermal process. Using standard tools of
statistical mechanics \cite{22} such as, for example, the
Kirkwood-Salsburg analysis, the correlation
inequalities of \cite{5}, homogeneous limits, we provide a class of
Euclidean invariant models of selfinteracting Bose matter that can be
controlled rigorously, as we shall demonstrate in section 3.

The unbounded (of polynomial type) perturbations of a free thermal
structure will be studied in another paper of this series \cite{18}. In
the critical region nonergodicity is preserved under gentle
perturbations (cf.\ the second part of \cite{18}), but whether this is
related to arising of the Bose condensate in an interacting system
remains to be resolved.

The pioneering paper of \cite{21} and the following ones \cite{23,24,25} have
provided, among others, the major inspirations for our own Euclidean
attitude to many bosons physics. The methods of classical statistical
mechanics have been
already applied to the studies of certain quantum systems in
\cite{24,26,27}, and, to
some extent, our approach to an interacting Bose gas resembles that of
the authors just quoted. %(is essentially based on similar ideas).

\setcounter{section}{1}
\sec
\section{Free Bose Gas(es). Euclidean Aspects}
The main aim of this section is to point out certain stochastic aspects
that arise in the Euclidean time of the thermal structure describing
systems of noninteracting Bose particles being in the thermal
equilibrium at (inverse) temperature $\beta >0$ and the chemical
activity $z$. Most of the results obtained below apply  well to the case
when the kinetic energy function $\Ep$ of the individual
particle is such that:
\bd
\i[$(i)$] $\forall_{t \in \bbR_+}e^{-t\Ep}$ is positive
definite continuous function of $p \in \bbR^d$ or equivalently that
\i[$(i')$] $\{e^{-t{\cal E}(-i\nabla )}, t\ge 0\}$ generates a
semigroup of positivity preserving operators on $L_2 (\bbR^d )$.
\ed
The most general form of such functions is given by the Levi-Khintchine
formula (see e.g. \cite{28,29})
\bel{Ep}
\Ep = a+i{\bf b}\cdot p+p\cdot C\cdot p -\int
[e^{ipx}-1-iph(x)]r(dx)
\ee
where: $a$ is some real constant, ${\bf b}$ is some vector in $\bbR^d$;
       $C$ is
       some nonnegative definite matrix  and $r$ is some nonnegative measure
       on $\bbR^d$, called the Levy measure, such that $\int_{\bbR^d} 1\wedge
       |x|^2 r(dx)<\infty$, where $x\wedge y\equiv min \{x,y\}$; $h$ is
       so called cut-off function with the compact support and
       satisfying $h(x) =x$ in some neighbourhood of the origin (see
       e.g.\cite{28} for explaining the role played by the cut-off function
       $h$ in this scheme).
In particular the functions $\Ep =|p|^{\alpha}; 0<\alpha\le 2$ or
$\Ep =\sqrt{p^2+m^2}$ belong to this class.
The common feature of all such functions is that the corresponding
semigroups $\{e^{-t{\cal E}(-i\nabla )},t\ge 0\}$ are generated by stochastic
Markov processes with stationary independent increments known as Levy
processes  \cite{28,29}.

The kernels of the semigroups $\{e^{-t{\cal E}(-i\nabla )},t\ge 0\}$ denoted as
${\K}^{({\cal E})}_t (x,y)$ have explicite expressions throughout
the corresponding path space integrals \cite{29}. This enables us to apply the
methods of \cite{1} to reproduce (up to some extent) the basic results of
\cite{1,2,3,4} for interacting gases with nonstandard
kinetic energy. The corresponding results shall be reported
elsewhere \cite{30}.

In the present paper we confine ourselves to the choice:\\
$\Ep =p^2$ called the standard Bose gas, and\\
$\Ep =\sqrt{p^2+m^2}, m\ge 0$ called the semirelativistic Bose
gas.

In the case of standard Bose gas the corresponding path space integral
is well known as Wiener (conditioned) integral and in this case the
corresponding transition function has a kernel
\beq
{\K}^s_t (x,y) =\frac{1}{(4\pi
t)^{d/2}}e^{-|x-y|^2/(4t)^{1/2}}
\eeq
In the case of semirelativistic Bose gas the corresponding transition
function has a kernel
\beq
{\K}^m_t (x,y)=\frac{4}{\pi^{1/2}}\int^{\infty}_0 d\tau {\cal
K}^s_{\tau}(x-y)\frac{e^{-t^2/4\tau } e^{-m^2 \tau}}{\tau^{3/2}}
\eeq
with fast exponential decay as $|x-y|\nearrow\infty$ for $m>0$ and in
the case $m=0$ equal to the well known symmetric Cauchy density:
\beq
{\K}^0_t (x,y) =\frac{c\cdot t}{(t^2+|x-y|^2)^{(d+1)/2}}
\eeq
\subsection{Global Aspects}
Let $\W(\gh )$be the abstract Weyl algebra built over the one-particle
space $\gh\equiv L_2(\bbR^d)$ equipped with the standard symplectic
form $\sigma (f,g)\equiv \Im\<f|g\>$. For a chosen kinetic energy function
$\Ep $ as above we define free thermal state
$\omega_0^{(\beta,\mu )}$ on the algebra ${\cal W}(\gh )$:
\beq
\omega_0^{(\beta,\mu )}(W(f))\equiv \exp{-\frac{1}{2}\int dp|\hat f
(p)|^2\hat C ^{\beta}_0 (p)}
\eeq
where:
\beq
\hat C ^{\beta}_0 (p) \equiv \frac{1+ze^{-\beta{\cal
E}(p)}}{1-ze^{-\beta\Ep }};
\eeq
$0<\beta$ is the (inverse) temperature, $z\equiv e^{-\beta\mu}$ is the
chemical activity and $\mu$ is the chemical potential. The values of $z$
 (corresponding to the noncritical regime
of the free Bose gas exclusively considered here) are restricted to:
$$
0<\sup\limits_{p}ze^{-\beta\Ep }<1
$$
which in the case $\Ep =p^2$ or $\Ep =|p|$ corresponds to
$0<z<1$ (resp. $(\mu >0 )$) and $0<ze^{-\beta m}<1$ (resp. $\mu >-m$) if
$m>0$ and $\Ep =\sqrt{p^2+m^2}$.

Some elementary properties of the free thermal kernel $C^{\beta}_0 (x)$
are collected in the following proposition.
\begin{prop}\label{p2.1}
For any noncritical value of $z$
the corresponding free thermal kernels $\Cbo (x)$ have the
following properties:
\bd
\i[(i)] $C^{\beta}_0(x)=\delta (x) +R^{\beta}_0 (x)$, where
$R^{\beta}_0 (x)>0$ for any $x\in \bbR^d$ and $R^{\beta}_0(x)\in
S(\bbR^d )$ if $\Ep =p^2$ or $\Ep =\sqrt{p^2+m^2}$ with
$m>0$.
\i[(ii)] if $\Ep =|p|$ then $C^{\beta}_0 (x)=\delta (x)
+R^{\beta}_0 (x)$ where $R^{\beta}_0 (x)>0$ and $R^{\beta}_0 \in C_0
(\bbR^d )\cap L_1 (\bbR^d )\cap C^{\infty}(\bbR^d)$.
\ed
\end{prop}
\proof
{}From the assumption $\sup\limits_{p} ze^{-\beta\Ep }<1$ we
obtain an equality:
\beq
\hat C^{\beta}_0 (p)=1+\hat R^{\beta}_0 (p)
\eeq
where $\hat R^{\beta}_0 (p)=2\sum^{\infty}_{n=1} z^n e^{-\beta n{\cal
E}(p)}$.\\
{}From the positive-definitness of the function $p\in \bbR^d
\longrightarrow e^{-t\Ep }$ for each $t>0$, it follows that for
each $n$, $\exp({-\beta n\Ep)}$ is the Fourier transform of some
positive measure $d\mu^{\beta}_n$ on $\bbR^d$. Moreover from the fact
that $\exp{-\beta n{\cal E}}(p) \in S(\bbR^d )$ in the case (i) it
follows that $d\mu^{\beta}_n (x) =\rho^{\beta}_n (x) d^d x$ , with
$\rho^{\beta}_n (x) \in S(\bbR^d )$. By elementary arguments it follows
that also $\sum^{\infty}_{n=1} z^n\exp{-\beta n {\cal E}}(p) \in S(\bbR^d )$
in the case (i), therefore we conclude that all assertions of (i) are
valid. The conclusions of (ii) follow from the explicite form (2.4) of
the corresponding kernels and elementary arguments.
\endproof

Let $({\cal H}_0,$ $\Omega_0, \pi_0 )$ be the corresponding GNS triplet
obtained from $({\cal W}(\gh )$, $\omega_0^{(\beta, \mu )})$. Then
defining $\alpha^0_t (\pi_0 (W(f))\equiv \pi_0
(W(z^{-it/\beta}e^{it\Ep }f))$ we obtain an $\sigma$-weakly
continuous group of automorphisms of $\pi_0 ({\cal W}(\gh ))''$. It
is well known that the system $\bbC_0 \equiv ({\cal
H}_0,\Omega_0,\alpha^0_t;\pi_0 ({\cal W}(\gh ))'')$ forms a $W^*$-KMS
system in the (inverse) temperature $\beta$ (see i.e. \cite{17}).
The corresponding multitime Green functions of the system $\bbC_0$ are
given by:
\beq
\begin{array}{rl}
G_0 ((t_1,f_1),\ldots ,(t_n,f_n ))\equiv&\omega^{(\beta ,\mu )}_0
(\alpha^0_{t_1}(\pi_0(W(f_1))\ldots \alpha^0_{t_n}(\pi_0(W(f_n)))=\\
=&\prod\limits_{1\le i\le j\le n}[\exp{i\sigma ((t_i ,f_i ),
(t_j,f_j))}\times\\
\null&\null\times\exp{-\frac{1}{2}\int \o{\^f_i (p)}\^f_j (p) \hat G_0^{\beta}
(t_i -t_j ;p) d{\bf p} }]
\end{array}
\eeq
where:
\beq
\sigma ((t_i, f_i ), (t_j, f_j ))=\Im<z^{-it_i /\beta}e^{it_i {\cal
E}(p)}\^f_i |z^{-it_j /\beta} e^{it_j \Ep}\^f_j >\,\,\,,
\eeq
\beq
\hat G^{\beta}_0 (t;p )\equiv \frac{z^{-it/\beta}e^{it{\cal
E}(p)}+z^{1+it/\beta}e^{-(\beta+it)\Ep }}{1-ze^{-\beta{\cal
E}(p)}}
\eeq
By elementary arguments they can be extended analitically to the
holomorphic functions $G_0 ((\zeta_1,f_1),\ldots ,(\zeta_n,f_n))$ of
${\bf \zeta}=(\zeta_1,\ldots ,\zeta_n ) \in T^{\beta}_n \equiv \{ {\bf
\zeta}^n =(\zeta_1, \ldots ,\zeta_n )\in \bbC^n |\ldots \Im \zeta_i <
\Im \zeta_{i+1} < \ldots ,\sum\limits^{n-1}_{i=1} (\Im \zeta_{i+1} -\Im
\zeta_i )<\beta \}$ and continuous on
$\overline{T}^{\beta}_n$. The restrictions of the analitically continued
Green functions to the so called Euclidean region $E^\beta_n
\equiv \{{\bf z}\in C^n |\Re z_i =0; -\beta /2 \le \Im z_1 \le \ldots
\le \Im z_i \le \Im z_{i+1} \le \ldots \le \beta /2\}$ will be
called Euclidean
Green functions of the free Bose gas and their full collection extended
to $\bigcup\limits_{n\ge 0}{\cal W}(\gh )^{\times n}$ by linearity
will be denoted by $^E \bbG^0$. The following
abbreviations will be used:
\beq
E^{\beta ,+}_n =\{(S_1,\ldots , S_n ) \in E^{\beta}_n |0\le S_i\};
\eeq
\beq
{\bf S}^k \equiv (S_1^k,\ldots S^k_k )\in E^{\beta}_k ;
\eeq
\beq
{\bf W}^k\equiv (W^k_1,\ldots ,W^k_k )\in {\cal W} (\gh )^{\times k};
\eeq
\beq
^E G^0_{{\bf W}^k}({\bf S}^k)\equiv ^E G^0_{W_1,\ldots ,W_k} (S_1,\ldots
S_k )
\eeq
\beq
{\bf f}^k \equiv (f_1, \ldots , f_k ) \in L_2 (\bbR^d )^{\times k};
\eeq
\beq
\ba{rcl}
^E G^0_{{\bf f}^k} ({\bf S}^k ) &\equiv& {}^E G^0_{f_1,\ldots , f_k
)}(S_1,\ldots ,S_k )\\
&=&  {}^E G^0_{W(f_1),\ldots , W(f_k))}(S_1,\ldots ,S_k )\\
\ea
\eeq
\beq
{\bf S}^{k*}\equiv (-S_k,\ldots ,-S_1 ) \quad \hbox{for} \quad {\bf S}^k
\in E^{\b}_k;
\eeq
\beq
{\bf W}^{k*}\equiv (W^+_k,\ldots , W^+_1 ) \quad \hbox{for} \quad {\bf
W}^k =(W_1, \ldots ,W_k );
\eeq
\beq
({\bf W}^n, {\bf S}^n ) \equiv ((W_1, S_1 ), \ldots , (W_n, S_n ))
\eeq
\begin{prop}\label{p2.2}
Let $^E \bbG^0=\{^E G_{W_1,\ldots ,W_k}(S_1,\ldots , S_k )|W_i \in {\cal
W}(\gh ), (S_1,\ldots , S_k )\in E^\b_k\}$ be the collection of the
Euclidean Green functions of the free Bose gas in the noncritical
regime. Then the collection $^E \bbG^0$ has the following properties:
\bd
\i[EG(1)]
\bd
\i[(i)] for each fixed ${\bf W}^k \in {\cal W}(\gh )^{\times k}$
the map\\ $E^{\beta}_k \ni {\bf S}^k \longrightarrow ^E G^0_{{\bf W}^k}
({\bf S}^k)$ \\is continuous
\i[(ii)] for each fixed ${\bf S}^k \in E^{\beta}_k$ the map\\
${\cal W}(\gh )^{\times k}\ni {\bf W}^k \longrightarrow ^E G^0_{{\bf
W}^k} ({\bf S}^k)$\\
is: multilinear and for any ${\bf f}^k \in L_2 (\bbR^d )^{\times k}$ the
map:\\
$L_2(\bbR^d )^{\times k}\ni {\bf f}^k \longrightarrow ^E G^0_{{\bf f}^k}
({\bf S}^k)$\\
is continuous and obeys the estimate $|^E G_{{\bf f}^k} ({\bf S}^k)|\le
1$.
\i[(iii)]for any ${\bf S}^k \in E^{\beta}_k$ and any $S\in [-\beta
/2, \beta /2]$ such that $S_k +S\le \beta /2$ the Euclidean Green
functions are locally shift invariant i.e. for any ${\bf W}^k \in {\cal
W}(\gh )^{\times k}$:\\
$^E G^0_{{\bf W}_k} ({\bf S}^k +S) =^E G^0_{{\bf W}^k} ({\bf S}^k)$\\
where ${\bf S}^k +S \equiv (S_1+S,\ldots ,S_k+S)$.
\i[(iv)] for any ${\bf W}^k \in {\cal W}(\gh )^{\times k}$, any
${\bf S}^k :\exists_{1\le i\le k-1}S_i=S_{i+1}$ we have the equality:
\\
$^E G^0_{{\bf W}^k}({\bf S}^k)=^E G^0_{{\bf W}^k_{(i)}} ({\bf S}^k_{(i)}$\\
where ${\bf W}^k_{(i)}=(W_1,\ldots ,W_{i-1},W_i \cdot W_{i+1},\ldots ,
W_k )\\
{\bf S}^k_{(i)}=(S_1,\ldots , S_i, S_{i+2},\ldots ,S_k )$
\i[(v)] for any ${\bf W}^k \in {\cal W}(\gh )^{\times
k}:\exists_{1\le i\le k}:W_i ={\bf 1}$ the following equality holds:
\\
$^E G^0_{{\bf W}^k}({\bf S}^k) =^E G^0_{_{(i)}{\bf W}^{k-1}}(_{(i)}{\bf
S}^{k-1})$\\
where \\
$_{(i)}{\bf W}^{(k-1)}=(W_1,\ldots , W_{i-1},W_{i+1}, \ldots ,W_k )$\\
$_{(i)}{\bf S}^{(k-1)}\equiv (S_1,\ldots ,S_{i-1}, S_{i+1}, \ldots S_k
)$
\i[(vi)] $^E G^0_{\bf 1}(0)=1$
\ed
\i[EG(2)] {\it (OS-positivity)}\\
For any terminating sequences
$$
\dtilde{\bf W}=({\bf W}^0,{\bf W}^1,\ldots {\bf W}^k,\ldots ),\quad
\dtilde{\bf S}=({\bf S}^0,\ldots ,{\bf S}^k,\ldots )
$$
with
$$
{\bf S}^k\in E^{\beta,+}_k \quad \hbox{for all} \quad k=1,2,\ldots
%\hbox{and for any} \quad f\in L_2(\bbR^d):
$$
\beq
\sum\limits_{k,l}{}^EG^0_{{\bf W}^{k *},{\bf W}^{l}}
({\bf S}^{k,*},{\bf S}^l) \geq 0\,,
\eeq
\i[EG(3)]
For any terminating sequences
$$
{{\bf W}}=({\bf W}^0,{\bf W}^1,\ldots {\bf W}^k,\ldots ),\quad
{\dtilde{\bf S}}=({\bf S}^0,\ldots ,{\bf S}^k,\ldots )
$$
with
$$
\dtilde{\bf S}^k\in E^{\beta,+}_k \quad \hbox{for all} \quad k=1,2,\ldots \quad
\hbox{and for any} \quad f\in L_2(\bbR^d):
$$
\beq
\sum\limits_{k,l}{}^EG^0_{\dtilde{\bf W}^{k,*},\overline{f};\~f,W^l}
({\bf S}^{k,*},0,0,S^l)
\leq \sum\limits_{k,l}{} ^EG^0_{\dtilde{\bf W}^k,\dtilde{\bf
W}^l}(\dtilde{\bf
S}^{k,*},\dtilde{\bf S}^l)
\eeq

\i[EG(4)] {\it (weak form of the KMS condition)}\\
Let $^E\hat{G}^0_{W_0,\ldots ,W_n}(S_1,\ldots ,S_n)\equiv
^EG^0_{W_0,W_1,\ldots ,W_n}(-\frac{\beta}{2},S_1-\frac{\beta}{2},\ldots
,S_n-\frac{\beta}{2})$ for $0\le S_1\le \ldots \le S_n\le \beta$. Then
for each $n,\quad {\bf W}^{n+1}\in{\cal W}(\gh )^{\times n}$
\beq
\ba{l}
^E\hat{G}^0_{{\bf W}^{n+1}}({\bf S}^n)=\\[1mm]
\dsq ^E\^G^0_{W_n,W_0,\ldots
,W_{n-1}}(\beta-S_n,\beta-S_n+S_1,\ldots ,\beta-S_n+S_{n-1})
\ea
\eeq
\i[EG(5)] {\it (Euclidean invariance and uniqueness of the vacuum)}\\
Under the natural action $\tau_{\{a,A\} }$ of the Euclidean Group of
Motions $E(d)$ on the Weyl algebra $W(\gh )$ the Euclidean Green
functions are
\bd
\i[(i)] invariant
\i[(ii)] have the cluster decomposition property, i.e.\\
for any ${\bf W}^k\in {\cal W}(\gh )^{\times k}, \quad {\bf W}^l \in
{\cal W}(\gh )^{\times l},\quad {\bf S}^k\in E^{\beta}_k,\quad {\bf
S}^l\in E^{\beta}_l$:
\beq
\lim\limits_{|a|\longrightarrow\infty} ^EG^0_{\tau_{\{a,0\}}{\bf
W}^k;{\bf W}^l}({\bf S}^k,{\bf S}^l)=^EG^0_{{\bf W}^k}({\bf
S}^k)\cdot^EG_{{\bf W}^l}({\bf S}^l)
\eeq
\ed
\ed
\end{prop}
\proof
Let us consider the free gas GNS $W^*$-KMS structure
  $\CC_0=(\H_0$, $\Omega_0$, $\alpha^0_t$, $\pi_0({\cal W}(\gh ))'')$. By the
Araki theorem \cite{31} the Euclidean Green functions are represented as:
\beq
^EG^0_{{\bf W}^n}({\bf S}^n)=\<\Omega_0|\alpha^0_{is_1}(\pi_0((W_1))\ldots
\alpha^0_{is_n}(\pi_0(W_n))\Omega_0\>
\eeq
and by the very definition of $\CC_0$:
\beq
\omega^{(\beta,\mu )}_0(W(f))=\<\Omega_0,\pi_0(W(f))\Omega_0\>
\eeq
Now everything follows easily from (2.24) and the Araki theorem.
In particular the OS positivity $EG(2)$ follows from the fact that the
$l.h.s.$ of (2.20) can be written as:
\beq
\begin{array}{ll}
\hbox{({\it l.h.s. of} (2.20))} \equiv&
\<\Omega_0|(\sum\limits_k\prod\limits_{l_k=1}^k \alpha^0_{is^k_{l_k}}
(\pi_0(W(f^k_{l_k}))))^+\right.\\
&\left. \cdot (\sum\limits_k
\prod\limits_{l_k=1}^k\alpha^0_{is_{l_k}^k}(\pi_0(W(f^k_{i_k})))\Omega_0\>
\end{array}
\eeq
The weak form of the KMS-condition, formulated as $EG(4)$ can be
observed easily from the explicite formulae (2.8) for the corresponding
Green functions.
\endproof
%\hfill{\begin{picture}(16,18)
%          \put(8,8){\circle*{8}}
%       \end{picture}
%      }\\

\remarks
As it was demonstrated in \cite{20} the multitime Euclidean Green
functions of any $C^*$--(or $W^*$)--KMS structure obey similar properties
$EG(1)\div EG(4)$ with the obvious modifications of the continuity
properties $EG(2)(ii)$ and $EG(3)$. It can be checked using the basic
results of \cite{1,2,3,4} that the Euclidean Green functions of Dilute Bose
gases (and also of Dilute Fermi gases built over the CAR algebra over
$\gh$) in the regime considered by Ginibre \cite{1} obey the system
$EG(1)\div EG(5)$. The detailed study of arising modular structures (see
below) are now under investigations. The Euclidean Green functions of
the critical Bose gas also obey properties similar to $EG(1)\div EG(5i)$
and their restrictions to the Abelian sector (of the Weyl algebra) fulfill also
$EG(6)$ (see below).
\endproof
The complex subalgebra ${\cal A}(\gh )$ of ${\cal W}(\gh )$
generated by the elements $W(f)$, with $f=\overline{f}$ will be called
an Abelian sector of ${\cal W}(\gh )$ and the corresponding free
Euclidean Green functions restricted to ${\cal A}(\gh )$ will be
denoted by $^{EA}\GG_0$.
%The explicite forms of $^{EA}\bbG_0$ are the following.
For $-\frac{\beta}{2}\le s_1\le \ldots\le s_n\le
\frac{\beta}{2}$ we have the following formulae:
\beq
^{EA}G_0((s_1,f_1),\ldots ,(s_n,f_n))=\prod\limits_{1\le i\le j\le
n}\exp{-\frac{1}{2}}S_0^{\beta}(s_j-s_i,f_i\otimes f_n)
\eeq
where:
\beq
S^{\beta}_0(s,f_i\otimes
f_n)=\int\hat{S}^{\beta}_0(s,p)\overline{f_i(-p)}f_j(p)\ dp ;
\eeq
\bel{2.28}
\hat{S}_0^{\beta}(s,p)\equiv \frac{z^{s/\beta}e^{-s{\cal
E}(p)}+z^{1-s/\beta}e^{-(\beta-s)\Ep }}{1-ze^{-\beta\Ep }}
\ee
The periodic extension of $\hat{S}^{\beta}_0(s,p)$ to the whole $\bbR$
shall be denoted by the same symbol. The fundamental properties of the
free thermal kernels $S^{\beta}_0(s,x)$ are collected in the following
Proposition.
\begin{prop}\label{p2.3} ~~~~\phantom{a}\\
\bd
\i[1.]
Let $\^S_0^\b$ be the free thermal kernel \r{2.28} with $\Ep=p^2$ or
$\Ep=\sqrt{p^2+m^2}$, $m>0$. Then for any $0\leq s\leq\b$; $z$ noncritical:
\bd
\i[(i)] $0<\Sbo(s,\void)\in S(\Rd)$ if $s \in (0,\ \b)$
\i[(ii)] $\Sbo(0,\void)=\Sbo(\b,\void)=\Cbo(\void)$ in $\DRd$ sense.
\ed
\i[2.] Let $\Ep=|p|$, then for any $0 \leq s \leq \b$, $0<z<1$
\bd
\i[(i)] $0<\Sbo(s,\void) \in L_1(\Rd)\cap C_0(\Rd) \cap C^\infty(\Rd)$
if $s \in (0,\ \b)$
\i[(ii)] $\Sbo(0,\void)=\Sbo(\b,\void)=\Cbo(\void)$ in $\DRd$ sense.
\ed
\i[3.] For $\Ep=p^2$ or $\Ep=\sqrt{p^2+m^2}$, $m\geq0$ the kernel
$\Sbo$ is
stochastically positive on the space $L_2(K_\b\times \Rd)$, i.e.\ for
any $g_1$, $g_2 \in L_2(K_\b)$; $f_1$, \ldots,$f_n\in L_2(\Rd)$, $C_1$,
\ldots, $C_n\in C$:
\bel{2.30}
\sum_{\a,\b=1}^n C_\a \o{C_\b} \Sbo({g_\a \tens f_\a}|g_\b \tens f_\b)
\geq 0
\ee
where
\bel{2.31}
\Sbo(g \tens f |g' \tens f') = \int_0^\b ds \int_0^\b ds'\,\o{g(s)}g'(s')
\int dx \int dy\,\o{f(x)} f'(y) \Sbo(|s-s'|,x-y)\,.
\ee
\i[4.]
 For $\Ep=p^2$ or $\Ep=\sqrt{p^2+m^2}$, $m\geq0$ the kernel $\Sbo$ is
 $OS$ positive on the circle $K_\b$, i.e.\ for any \n t $in[0,\b/2]$,
 \n f $\in L_2^r(\Rd)$, \n c $\in C$:
 \bel{2.32}
 \sum_{\a,\b}\o c_\a c_\b\int dx \int dy\, \Sbo(t_\a+t_\b|f_\a \tens f_\b) \geq
0
 \ee
 \ed
 \end{prop}
\proof
{}From the assumption $\sup\limits_{p} ze^{-\beta\Ep }<1$ it follows
that:
\bel{2.33}
\^\Sbo(s,p)=\sum_{n\geq 0} \^{F_n}(s,p)
\ee
where:
\bel{2.34}
\^F_n(s,p)\equiv z^{n+s/\b}e^{-(\b
n+s)\Ep}+z^{n+1-s/\b}e^{-(\b(n+1)-s)\Ep}
\ee
So, if  $\Ep=p^2$ or $\Ep=\sqrt{p^2+m^2}$, $m>0$ then
$\^F_n(s,p)\in S(\Rd)$ for each $n \geq 1$ and $ n=0$ if $s \in (0,\
\b)$. In this case also $\dsp \sum_{n \geq 1}^\infty z^n \exp{(-\b n
\Ep)} \in S(\Rd)$.
Taking into account that
$$
\^{\Sbo}(s,p)=\(\sum_{n \geq 0} z^n e^{-\b n \Ep} \)
\(z^{s/\b}e^{-s\Ep}+z^{1-s/\b} e^{-(\b-s)\Ep}\)
$$
it follows that also $\^{\Sbo}(s,p)\in S(\Rd)$ if $s\in(0,\ \b)$.
Moreover
$\Sbo(s,x)>0$ for any $x \in \Rd$.

Similarly, if $\Ep=|p|$ than we have:
\bel{2.35}
\^\Sbo(s,p)=\(\sum_{n\geq 0} z^n e^{-\b n |p|}\)
\(z^{s/\b}e^{-s|p|}+z^{1-s/\b}e^{-(\b-s)|p|}\)
\ee
%is a continuous function of $p$ for any $s\in [0,\b]$ and the series in
%\r{2.35} is convergent in $L_2(\Rd)$ whenever $s\in (0,\b)$.
therefore from the
continuity of Fourier transform and (2.4) we obtain
\bel{2.36}
\ba{rcl}
\dsp \Sbo(s,x)&=&\dsp \sumno{z^n c \over (\b^2n^2+|x|^2)^\dh} \\
&&\dsp \({z^{s/\b}c \over (s^2+|x|^2)^\dh}\right.
 +\left.{z^{1- s/\b}c \over (\b-s)^2+|x|^2)^\dh}\)\\
&=&\dsp \sumno z^{n+s/\b}{c \over \((\b n +s)^2+|x|^2\)^\dh}\\
&& \dsp+
\sumno z^{n+1-s/\b}{c \over \((\b (n+1) +s)^2+|x|^2\)^\dh}\,.
\ea
\ee
Because the above series are uniformly convergent on
$\Rd$  and define a continuous function with decay at
least as $\dsp {1 \over (s^2+|x|^2)^\dh}$ for $|x| \uparrow \infty$,
which is integrable provided $s>0$.

Although the claims 3 and 4 follow easily from a basic characterization
theorem of KL \cite{32} we present simple proofs of them for reader's
convenience. Expanding into the Fourier series the periodic function
$\^S^\b(s,p)$ we obtain:
\bel{2.37}
\ba{rcl}
\dsp \^S^\b(s,p)&=&\dsp \sum_{n \in \ZZ} \((\mu\b+\Ep)^2+(2\pi
n)^2\)^{-1}\(2\b(\mu\b+\Ep\)\\
&\cdot&\dsp \(1-e^{-\b(\mu+\Ep)}\)\(1-z\,e^{-\b\Ep}\)^{-1}
e^{i2\pi n s /\b}\,.
\ea
\ee
Because all the Fourier coefficients in the expansion
\r{2.37} are positive the stochastic positivity \r{2.30} follows. The OS
positivity of one-time Euclidean Green function is a general feature of
all KMS systems as was demonstrated in \cite{19}. The straightforward
proof of 4) is as follows. Let
$$\CC_0=(\H_0,\Om_0,\pi_0,\a^0_t;\ \pi_0(\W(h))'')$$ be the basic GNS
$W^*$-KMS system of the free Bose gas. Then we can write:
\bel{2.38}
\ba{rcl}
\dsp \sum_{\a,\b}c_\a c_\b\Sbo(s_\a+ s_\b | \o f_\a \tens f_\b)
&=&\dsp ||\sum_\a c_\a \a^0_{is_\a}(\pi_0(\W(f_\a))\Om_0||^2\\
&\geq&0
\ea
\ee
\endproof
\remarks
\ben
\i Let $h^\mu_0$ be a nonnegative, selfadjoint generator of unitary
group $U^0_t=z^{-it/\b}e^{-it\Ep}$ acting in the space $\gh=L_2(\Rd)$
and let $dP^\mu$ be the corresponding spectral measure of $h^\mu_0$.
Then defining the covariance operator
\bel{2.39}
\Gbo(s)\equiv \intR { d P^\mu(\lam) \over 1 -
e^{-\b\lam}}\(e^{-s\lam}+e^{-(\b-s)\lam}\)
\ee
acting in $\gh$ by definition:
\bel{2.40}
\<f,\,\Gbo g\> \equiv \^\Sbo(s|f \tens g)
\ee
we see that the kernel $\Sbo(s|f \tens g)$ belongs to the class of
kernels considered in \cite{32}.

\i Let us observe that the periodic kernels $\nSb(s,p) \equiv
F_n(s,p)$ for each $n$ also have the positivity properties stated in the
points 3) and 4) of proposition \ref{p2.3}. This leads to an interesting
decomposition of the free thermal process $\xi^0_t$ defined below as a
sum of independent OS-positive Gaussian processes  $\xi^{0,n}_t$, which
have covariances equal to $\nSb(s,p)$. This decomposition might be
eventually used to develop a rigorous Renormalisation Group Analysis of
interacting Boses Gases.
\een

\begin{prop}\label{p2.4}
The collection $^{EA}\GG_0$ of the Euclidean Green functions of the free
Bose Gas in the noncritical regime obeys the properties EG(1)--EG(5) of
proposition \ref{p2.2} and additionally:
\bd
\i[EG(6):] {\it (stochastic positivity)}\\
for any: $\dtilde S^k \in E^\b_k$, $f^k=(f^k_1$, \ldots, $f^k_k)$:
$f^k_i=\o f^k_i \in L_2(\Rd)$,
\bel{2.41}
 \sum_{k,l}\,{}^E G^0_{\dtilde f^{k*},\,\dtilde f^l}(\dtilde s^k,
\, \dtilde s^l) \geq 0
\ee
\ed
\end{prop}
\proof
%It follows easily from Proposition \ref{p2.0} below remarking that the
%r.h.s of \r{2.4} is equal to
%$$ \EE \left| \sum_k \prod_{l_k=1}^k \exp i
%\<\xi^0_{S^k_{l_k}};\,f^k_{l_k}\>\right|^2\,,
%$$
%where $\dsp \xi^0_t$ is the free thermal process defined below.
{}From assertion 3 of Prop \ref{p2.3} it follows by standard construction
(see i.e. \cite{28,32}) that there exists a Gaussian process
$(\xi^0_t)_{t \e K_\b}$ indexed by $L^2(\Rd)$, with mean zero and the
covariance given by $S_0^\b (\tau,\ x)$. The r.h.s of \r{2.41} can be
rewritten in terms of $(\xi^0_t)$ as
$$
\EE \left| \sum_k \prod_{l_k=1}^{n_k} \exp i
\<\xi^0_{s^k_{l_k}};\,f^k_{l_k}\>\right|^2\,,
$$
\endproof

Having defined a system of Euclidean multitime Green functions with the
properties listed in Proposition \ref{p2.2} we can apply the
constructions of \cite{20} to build certain $W^*$-KMS structures. The
interesting aspect of the proposition below is that the system of
Euclidean Green functions of the free Bose Gas restricted to ${\cal A}(\gh)$
already contains all information of the free Bose Gas.
\begin{prop}\label{p2.5}
Let $\Ep=p^2$ or $\Ep=\sqrt{p^2+m^2}$, $m \geq 0$ and let $z$ be
noncritical. Then
\ben
\i There exists a unique (up to a unitary equivalence) $W^*$-KMS system
$^E\CC=({}^E\H_0,\,{}^E \Om_0,\,{}^E \a^0_t,\,{}^E \m_0)$ and a bounded
$^*$-representation $^E \pi_0$ of $\Wh$ such that:
\ben
\i[(i)] $^E\pi_0(\Wh) \subseteq ^E \m_0\,$,
\i[(ii)] the multitime Euclidean Green functions of $^E \CC_0$
restricted to $^E\pi_0(\Wh)$ coincide with $^E\GG_0$.
\i[(iii)]
$$
\E\mo = W^*\{ \E\aot1(\E\pi_0(W(f_1)))\ldots \E\aot n (\E\pi_0(W(f_n)))\}\,,
$$
\een
\i  There exists a unique (up to a unitary equivalence) $W^*$-KMS system
$^A\CC=({}^A\H_0,\,{}^A \Om_0,\,{}^A \a^0_t,\,{}^A \m_0)$ and a bounded
$^*$-representation $^A \pi_0$ of ${\cal A}(\gh)$ such that:
\ben
\i[(i)] $^A\pi_0({\cal A}(\gh)) \subseteq ^A \m_0\,$,
\i[(ii)] the multitime Euclidean Green functions of
the system $\A\CC_0$ restricted to $^A\pi_0({\cal A}(\gh))$ coincide with
${}^A \GG_0$.
\i[(iii)]
$$
\A\mo = W^*\{ \A\aot1(\A\pi_0(W_1))\ldots \A\aot n
(\A\pi_0(W_n))\}\,,$$
for $W_1,\ldots,W_n \in{\cal A}(\gh)\,.$
\een
\i Both systems $\E\CC_0$ and $\A\CC_0$ are unitarly equivalent to the
GNS $W^*$ KMS system $\CC_0=(\Ho,\Omo,\aot{},\pi_0(\Wh)'')\,$.
\een
\end{prop}

\proof
{\bf Step 1.}\\
In the first step we apply in a sketchy way a general construction of
\cite{20} (see also \cite{19}) to which we are referring for more
details. Because in both cases the constructions of $\E\CC_0$ and
$\A\CC_0$ are identical we shall restrict ourselves to the construction
of $\E\CC_0$ only.

Let $\~V^\b$ be the free complex vector space built over the set
$\{(\dW^n,\,\ds^n)|\,\ds^n \in E_n^{\b,+}\}$. Then we divide $\~V^\b$ by
the natural relations arising from the properties EG(1)(i), EG(1)(iv),
EG(1)(v)  and EG(1)(vi) obtaining a complex vector space $V^\b$. The
following sesquilinear form
\bel{2.42}
\(\sum_\a c_\a \dWs{n_\a} ; \sum_\b d_\b \dWs{k_\b}\)
\equiv
\sum_{\a,\b} \o c_\a d_\b
\E G^o_{\dW^{n_\a *},\dW^{k_\b}}
\(\ds^{n_\a *} ,\,\ds^{k_\b}\)
\ee
defined on $V^\b$ is nonnegative by EG(2). The corresponding Hilbert
space will be denoted by $\E\Ho$ and the
corresponding classes of abstraction  will be denoted by square
brackets "[ ]".

Lifting the natural action $\E\~\pi_0$ of $\Wh$ on $\~V^\b$, defined by
$$ \E\~\pi_0(W)\dWs n \equiv ((W,\dW^n);\,(0,\ds^n))\,,$$
to the space $\E\Ho$ we obtain a $^*$-representation of $\Wh$ in $\E\Ho$
which is bounded because of EG(3).

Lifting the local shift transformation given by EG(1)(iii) into the
space $\E\Ho$ we obtain a uniquely determined selfadjoint generator $\E
H_0$. Defining $\E\Omo=[(1,0)] \in \E\Ho$ we have that for any $[\dWs n]
\in \E\Ho$:
$$
\E\a^0_{is_n}(\Epo(W_1))\,\ldots \E\a^0_{is_1}(\Epo(W_n)) \E\Omo =[\dWs
n]\,.
$$
Moreover, the vector valued maps:
$$
E^{\b,+}_n \ni \ds ^n \to \prod_{k=1}^n \E\a^0_{is_k}(\Epo(W_k))\E\Om
\in \cal H^E
$$
can be holomorphically extended to the tube $T^n_\b$ being continuous on
the boundary $\p T^n_\b$. In particular it can be proved (see \cite{20})
that the
vector $\E\Omo$ is cyclic and separating for the $W^*$-closure $\E\m_0$
of the $^*$-algebra generated by all products: $\E\aot
1(\Epo(W_1))\ldots \E\aot n (W_n))$ where $t_1,\,\ldots,t_n\in\RR$;
$W_1,\, \ldots,W_n \in \Wh$. Thus we have sketched the construction and
the proof that $\E\CC_0 \equiv (\E\Ho,\E\Omo;\E\aot{};\E\m_0)$ forms a
$W^*$ KMS system. The Euclidean Green functions of the system $\E\CC_0$
are equal to $\E\GG_0$ by the very construction.
Let $\E\CC_0' \equiv (\E\Ho',\E\Omo';\E\aot{}{}';\E\m_0')$ be another
$W^*$ KMS system whose the Euclidean Green function coincide with
$\E\GG_0$ and such that $\E\m_0 \supset \Epo'(\Wh)$ for some $\Epo' \in
\fn{Rep}^*(\Wh,L(\E\Ho'))$ and
$\E\m_0'=W^*\{\E\aot1{}'(\Epo'(W_1))\,\ldots\E\aot n{}'(\Epo'(W_n))\}$.
Then the isometry
$$
\ba{rcl}
\dsp j&\,:&\dsp\,\, \E\aot1(\Epo(W_1))\,\ldots\E\aot n(\Epo(W_n)) \E\Omo\\
&& \E\aot1{}'(\Epo'(W_1))\,\ldots\E\aot n{}'(\Epo'(W_n)) \E\Omo'
\ea
$$
can be extended to a unitary operator  such that $j\E\Omo=\E\Omo'$;
$\E\aot{}= j^{-1} \E\aot{}{}'j$; $\E\m_0'=j\E\m_0 j$.

{\bf Step 2.}\\
In the second step we identify the $W^*$-KMS system $\E\CC_0$ with
$\CC_0$. Although this identification follows from Section V of \cite{20} we
present straightforward proof below. To start with let us define a
linear space ${\cal D}^E$ generated by:
$$\ba{r}
\dsp \{\E\a^0_{is_n}(\Epo(W(f_n)))\ldots \E\a^0_{is_1}
(\Epo(W(f_1)))\E\Omo | \\
\dsp \ds^n \in E^{\b,+}_n\,,\,\dtilde f^n \in L^2(\Rd)^{\otimes n}\}
\ea
$$
{}From the step 1.\ we know that $\o{{\cal D}^E}=\E\Ho$ and for any $\dtilde
f^n
\in L_2(\Rd)^{\x n}$ the map
$$
E^{\b,+}_n \ni \ds^n \to T- \prod_{i=1}^n \E\a^0_{is_i} (W(f_i)) \E\Omo
$$
can be uniquely extended to a holomorphic, vector valued function on the
tube $T^\b_n$ and this extension gives also the holomorphic extension of
the corresponding Green function,

Computing the r.h.s.\ of
\bel{2.43}
\ba{l}
\dsp \<\apW {\tau_n}{f_1}{\tau_1}{f_n} \E\Omo,\right.\\
\dsp \left.\E\aot{}(W(f)) \apW{s_m}{g_1}{s_1}{g_m} \E\Omo\> \\
\quad \dsp = G^0 \((\o f_n,\,-i\tau_n),\,\ldots,(\o f_1,\,-i\tau_1);
\,(f,t),\,(g_1,is_1),\,\ldots,(g_m, is_m)\)
\ea
\ee
with the help of the formulae (2.8)
and comparing it with
\bel{2.44}
\ba{l}
\dsp \<\apW{\tau_n}{f_n}{\tau_1}{f_1}\E\Omo;\right.\\
\quad \dsp \,
W[z^{-it/\b}e^{it\Ep}f]\cdot\left.\apW{s_n}{g_n}{s_1}{g_1}\E\Omo\>
\ea
\ee
we conclude that:
\bel{2.45}
\E\a^0_t\(\Epo(W(f))\) = \Epo\(W(z^{it/\b}e^{it\cal E}f)\)
\ee
on a dense domain ${\cal D}^E$ and thus on $\E\Ho$.

Defining a map,
\bel{2.46}
\ba{l}
\dsp j_E\,:\,\, \apW{s_n}{f_n}{s_1}{f_1}\E\Omo \\
\dsp \qquad \to \a^0_{is_n}(\pi_0(W(f_n)))\,\ldots \a^0_{is_1}(\pi_0(W(f_n)))
\Omo \in \Ho
\ea
\ee
we obtain a densely defined with a dense range isometry from $\E\Ho$ to
$\Ho$ which extends naturally to a unitary map $j_E$. From \r{2.45} we
have:
$$ j_E{}\,^E\aot{}j\1 = \aot{}\,,\quad j_E{}\,^E\Omo = \Omo \quad \mbox{ and }
\quad \E\m_0=j_E\1 \pi_0(W(h))\2 j_E\,.$$

{\bf Step 3.}\\
In the third step we identify the $W^*$-KMS system $^{A}\CC_0$ with
$\CC_0$. The following lemma, whose proof is translated into the fourth
step below plays a basic role.
\begin{lem}\label{l2.6}
Let $\Ep=p^2$ or $\Ep=\sqrt{p^2+m^2}$, $m\geq 0$. Then the set of
functions
$$
V=\{ e^{it\Ep}f(p) |\,t \in \RR;\, f=\o f \in L_2(\Rd)\}
$$
is $\RR$-linearly dense in $L^2(\Rd)$.
\end{lem}

It is because the Euclidean Green functions restricted to the abelian
sector ${\cal A}(\gh)$ of $\Wh$ obey the properties EG(1)--EG(4) we can
apply the construction presented in the step 1.\ obtaining again a
$W^*$-KMS system $\A\CC_0=(\A\Ho,\A\Omo,\A\aot{};\,\A\m_0)$ where
$\A\m_0$ is the $W^*$-algebra generated by the operators
$\A\aot1(\A\pi_0(W(f_1)))\,\ldots\A\aot n(\A\pi_0(W(f_n)))$, where
$\A\pi_0$ is the corresponding representation of $\Ah$ in $L(\A\Ho)$ and
all $f_i$ are real. From the cyclicity of $\A\Omo$ under the action of
$\A\m_0$ it follows that the set of vectors
$$ \aPWbi{t_1}{f1}{t_n}{f_n}{\A} \A\Omo
$$
is linearly dense in $\A\Ho$. Defining a map
\bel{2.47}
\ba{l}
\dsp j_A \: \aPWbi{t_1}{f_1}{t_n}{f_n}{\A} \A\Omo\\
\dsp \quad \to \aPWbi{t_1}{f_1}{t_n}{f_n}{} \Omo \\
\dsp \qquad \equiv \pi_0 \(W(\sum_{\a=1}^n e^{it_\a h^\mu}f_\a)\)\Omo
 \prod_{1 \leq \a < \b \leq n}
\exp \{-i\sigma \(e^{it_\a h^\mu}f_\a;\,e^{it_\b h^\mu}f_\b\)\}
\ea
\ee
we see that it is an isometry with dense range because of Lemma
\ref{l2.6}. Moreover $j_a({}^A\Om_0)=\Om_0$.

Computing
\bel{2.48}
\ba{l}
\dsp \(j_A \sum_{l=1}^n \apw{t_l}{g_l}{\A} j_A^*\)
\prod_{k=1}^m \pi_0 (W(e^{-i(S_k h^\mu)}f_k))\Omo \\
%\quad\dsp  = j_a \left[\(\sum_{l=1}^n \apw{t_l}{g_l}{\A}\)
%        \(\prod_{k=1}^m\apw{s_k}{f_k}{\A}\) \A\Omo \right] \\
\quad \dsp = \prod_{l=1}^n\apw{t_l}{g_l}{}\,
             \prod_{k=1}^m\apw{s_k}{f_k}{}\Omo
\ea
\ee
we obtain
\bel{2.49}
j_A\( \prod_{k=1}^m\apw{t_k}{g_k}{\A}\)j_A^* =
\prod_{k=1}^m \apw{t_k}{g_k}{}\,,
\ee
therefore applying Lemma \ref{l2.6} again we conclude that:
\bel{2.50}
j_A\(\A\m_0\)j_A^* =\pi_0 (\Wh)\2
\ee

Let us observe also that the map
\bel{2.51}
\A\~\pi_0\: W\(\sum_\a e^{it_a h\mu}f_\a\)
        \to
\prod_{1 \leq \a <\b \leq n} \exp\{i\sigma \(e^{it_\a h} f_\a,\,e^{it_\b
h} f_\b\) \prod _\a \apwbi{t_\a}{f_\a}{\A}
\ee
can  be extended  to  representation of the full Weyl algebra
$\Wh$ in $L(\A\Ho)$ and moreover the obtained representation extends
$\A\pi_0'$. For this, let us observe that:
\bel{2.52}
\ba{l}
\dsp \A\~\pi_0\(W(\sum_\a e^{it_\a h^\mu}f_\a)\) \\
\quad \dsp =
\prod_{1 \leq \a <\b \leq n} \exp\{i\sigma \(e^{it_\a h^\mu} f_\a,\,e^{it_\b
h^\mu} f_\b\)\} \prod _\g \apwbi{t_\g}{f_\g}{\A} \\
\quad \dsp =
\prod_{1 \leq \a <\b \leq n} \exp\{i\sigma \(e^{it_\a h^\mu} f_\a,\,e^{it_\b
h^\mu} f_\b\)\} j_A\1\( \prod _\g \apwbi{t_\g}{f_\g}{}\) j_A \\
\quad \dsp = j_A\1 \(\pi_0(W(\sum_\g e^{it_\g h^\mu}f_\g))\) j_A\,,
\ea
\ee
by using \r{2.49},        and the fact that $\pi_0$ is a representation of
$\Wh$. From lemma \ref{l2.6} we know that for any $g \in L^r_2(\Rd)$ there
exists a sequence
$$
(t^\a_1\ooo t^\a_{n_\a}),\,(f^\a_1\ooo f^\a_{n_\a}) \in L_2(\Rd)
$$
such that $\sum_k \exp\{it^\a_kh^\mu\}f^\a_k\to g$ in $L^2(\Rd)$ sense.

Because $\pi_0$ is $L_2(\Rd)$ continuous representation of $\Wh$ it follows
that the limit
$$
\lim_{\a \rightarrow \infty} j_A\1\(\pi_0\(W(\sum_k
e^{it^\a_kh^\mu}f^\a_k)\)\)j_A
$$
exists in the weak sense, therefore we conclude that also
\bel{2.53}
\lim_{\a \rightarrow \infty}
\prod_{1 \leq k_\a <l_\a \leq n} \exp\{i\sigma \(e^{it^\a_{k_\a} h^\mu}
f^\a_{k_\a},\,e^{it^\a_{l_\a}h^\mu} f^\a_{l_\a}\)\}
\prod _\g \apw{t_\g}{f_\g}{\A}
\equiv \A\~\pi(W(g))
\ee
exists in the weak sense. Now it is easy to check that $\A\~\pi$ as
defined in \r{2.53} really *-bounded representation of $\Wh$ in $L(\A\Ho)$
and such that $\A\~\pi_0{}_{|\Ah} = \A\pi_0$.
\endproof

{\bf Step 4. proof of Lemma \ref{l2.6}.}\\

The operator $e^{it\D}$ acts as $e^{it\D} f= \(e^{itp^2} \hat
f\)\check{}$,
where $\hat{}$ and $\check{}$ denote the Fourier transform and its
inverse. Let us take $g\in C_c(\Rd)$ which is a dense subspace in
$\LRd$. Let $g_1(p) = \half[g(p)+\o{g(-p)}]$ and
$g_2(p) = {1 \over 2i}[g(p)-\o{g(-p)}]$ be hermitian parts of $g$.
Because $g_1$ is the Fourier transform of a real valued function we may
write
\bel{2.54}
\ba{l}
\dsp \| g(p)-[\sum_{k=1}^n\hat f_k(p) e^{it_kp^2}- \sum_{k=1}^n\hat f_k(p)
e^{it_kp^2}
-g_1(p)]\|_{L^2} \\
\quad \dsp = \| ig_2(p) - i \sum_{k=1}^n\hat f_k(p) \sin(t_kp^2)\|_{L^2}
\ea
\ee
so it is enough to show that for every $\ve>0$ there exist real
valued functions $f_1\ooo f_n \in \LRd$ and $t_1\ooo t_n \in \RR$ such
that
\bel{2.55}
 \| g_2(p) -  \sum_{k=1}^n\hat f_k(p) \sin(t_kp^2)\|_{L^2} < \ve\,.
 \ee
Let $B$ denote a ball in $\Rd$ of radius $c>0$ such that $\fn{supp} g(p)
\subset B$.
Let $\hat f_k(p) = a_k(p)g_2(p)$, where $a_k(p)\in C_0(\Rd) \cap \LRd$
and $ a_k(p)= \o{a_k(p)}=a_k(|p|)$.
It is clear that $\hat f_k(p)$ is hermitian and belongs to $\LRd$. Then
\bel{2.56}
\ba{l}
\dsp \|g_2(p) - \sum_{k=1}^n \hat f_k(p) \sin(t_kp^2)\|_{L^2}\\
\quad \dsp \leq \|g_2\|_\infty \|1  - \sum_{k=1}^n a_k(p)
\sin(t_kp^2)\|_{L^2(B)}\,.
\ea
\ee
Let us deform the constant function $1$ to a function $f_0 \in C(B)$
such that $f_0(p)\geq 0\,\, \forall_{p\in B}$, $f_0(0)=0$ and
$\|1-f_0\|_{l^2(B)} < \ve$, $f_0(p)=f_0(p')$ if $|p|=|p'|$.
Then
\bel{2.57}
\ba{l}
\dsp \|1-  \sum_{k=1}^n a_k(p) \sin(t_kp^2)\|_{L^2(B)}\\
\dsp \quad \leq \ve +\mu(B)^\half
\sup_{|p| \in [0,c]} | f_0(|p|) - \sum_{k=1}^n a_k(|p|)
\sin(t_kp^2)|\,,
\ea
\ee
where $\mu(B)$ is the Lebesque measure of the ball $B$.

%Now we show that finite sums $ \sum_{k=1}^n a_k(p)
%\sin(t_kp^2)$ form a real subalgebra of $C_0(\BmO)$, where
%$a_k(p)$ and $\sin(t_kp^2)$ are considered as restricted to $\BmO$.
%It is enough to check that product of $ a_1(p) \sin(t_1p^2) $
%and $ a_2(p) \sin(t_2p^2) $ has an appropriate form. Let
%$a(p)=a_1(p)a_2(p)\sin(t_1p_2)$. Then $a(p)$ is real valued, depends only
%on $|p|$ and belongs to $C_0(\Rd) \cap \LRd$.
%Because
%\bel{2.58}
%\ba{l}
%\dsp \sup_{p \in \BmO} |f_0(p) - \sum_{k=1}^n a_k(p) \sin(t_kp^2)| \\
%\quad \dsp = \sup_{|p|\in[0,c]} |f_0(|p|) - \sum_{k=1}^n a_k(|p|)
%\sin(t_k\,|p|^2)
%\ea
%\ee
We consider a real algebra generated by
$ \sum_{k=1}^n a_k(|p|) \sin(t_k\,|p|^2)$ on $(0,c]$.
It is clear that $\sin(t\,|p|^2)$ separates points in $(0,c]$ and for every
$|p|\in %))
(0,c]$ there exists $t\in \RR$ such that $\sin t |p|^2 \neq 0$. %)
Because we may choose $a(p)$ such that $\a_{|B}=1$ so our algebra
separates points and nowhere vanishes in $(0,c]$. Thus applying  %[)
Stone-Weierstrass theorem to $C_0(0,c]$ we have that  %)
$$
\dsp  \sup_{|p|\in(0,c]} | f_0(|p|) - \sum_{k=1}^n a_k(|p|) %)
\sin(t_k\,|p|^2)| < \ve
$$
for some $a_1\ooo a_n$ and $t_1 \ooo t_n$. Finally
$$
\| g_2(p) -  \sum_{k=1}^n\hat f_k(p) \sin(t_kp^2)\|_{L^2}
\leq \|g_2\|_{\fn{sup}}(1+\mu(B)^\half)\ve
$$ which proves the assertion for $\Ep=p^2$. The same proof works for
$\Ep=\sqrt{p^2+m^2}$, $m\geq0$.
\endproof

To exploit the stochastic positivity EG(6) of the system $^{AE}\GG$ and
for the further development we shall introduce two basic concepts of
the generalized thermal process and the generalized thermal random
field.

If should be emphasized that these concepts are heavily inspired by the
abstract theory developed by Klein and Landau in \cite{25} (see also
\cite{32}).

\begin{df}\label{d2.7}
Any generalized, periodic (with the period $\b$) stochastic process
$(\xi_t)_{t \in \RR}$ with values in $\DRd$ will be called a
thermal process (with the temperature $\b$) iff
\bd
\i[Tp(1)] the process $(\xi_t)_{t \in \RR}$ is symmetric on $K_\b$, i.e.
\bel{2.59}
\forall_{-\b/2 \leq \tau \leq \b/2} \forall_{f \in {\cal D}(\Rd)}
\, \<\xi_\tau,\,f\> = \< \xi_{-\tau},\, f\> \qquad\mbox{(in law)}
\ee
\i[Tp(2)] the process $(\xi_t)_{t \in \RR}$ is (locally) homogeneous
i.e.
\bel{2.60}
\forall_{\tau,s\in K_\b \atop \tau+s \leq \b/2}
\forall_{f\in {\cal D}(\Rd)}
\, \<\xi_{\tau+s},\,f\> = \< \xi_{\tau},\, f\> \qquad\mbox{(in law)}
\ee
\i[Tp(3)] the process $(\xi_t)_{t \in \RR}$ is OS-positive on $K_\b$
i.e.\ for any bounded $F\in C_b(\RR^n)$, any $\dt^n \in [0,\b/2]^{\x
n}$; $\dtilde f^n \in {\cal D}(\Rd)^{\x n}$
\bel{2.61}
0 \leq E F\(\xf{-\tau_1^n}{1}\ooo\xf{-\tau_n^n}{n}\)
      F\(\xf{\tau_1^n}{1}\ooo\xf{\tau_n^n}{n}\)
\ee
\i[Tp(4)] the moments:
\bel{mom}
E\(\prod_{i=1}^n e^{i\xf{\tau_i}i}\) \equiv G_{f_1\ooo f_n}^{(\xi)}
(\tau_1\ooo\tau_n)
\ee
are continuous in $\dt^n \in (K_\b)^{\x n}$ and on $\Drd\xn$
\ed
A thermal process $\xi$ is called Euclidean invariant if additionally:
\bd
\i[Tp(5)] the moments \r{mom} are invariant under the action of the
Euclidean Group $E(d)$ in ${\cal D}(\Rd)$.
\ed
A thermal process $(\xi_t)_{t \in \RR}$ is called tempered iff the
moments \r{mom} are continuous on $S(\Rd)\xn$, $L_p$-continuous iff the
moments \r{mom} are continuous on $L^p(\Rd)\xn$, etc.
\end{df}

If $(\xi_t)_t$ is a generalized thermal process then its corresponding
path space measure construction leads to to concept of the random
generalized field.
\begin{df}\label{d2.8}
Any generalized random field $\mu^\b$ on ${\cal D}'(K_\b\x\Rd)$ (i.e. any
probabilistic, Borel cylindric (PBC) measure) will be called generalized
thermal random field iff
\bd
\i[Tf(1)]
\bel{2.62}
\forall_{g \in C^\infty(K_\b) \atop f\in\Drd}
\< \phi;\,r(g\tens f)\>=\< \phi;\,g\tens f\>
\quad \mbox{(in law $\mu^\b$)}
\ee
where $  r(g\tens f)(\tau,x)=g(-\tau)f(x)\,$,
\i[Tf(2)]
\bel{2.63}
\forall_{g \in C_0^\infty(K_\b) \atop f\in\Drd}
\< \phi;\,t_s(g\tens f)\>=\< \phi;\,g\tens f\>
\quad \mbox{(in law $\mu^\b$)}
\ee
for any $s>0$ such that $\supp t_s(g) \subseteq [-\b/2,\,\b/2]$, where
$t_s(g)(\tau) \equiv g(t+s)\,$,
\i[Tf(3)] the field $\mu^\b$ inOS-positive on the circle $K_\b$, i.e. for any
bounded cylindric function $F$ based on $(g_1\tens f_1\ooo g_n\tens
f_n)$, where $g_i\in C^\infty[0,\b/2]$ for all $i$, $f_i\in\Drd$:
\bel{2.64}
0\leq \mu^\b\(RF\(\fgf1{}\ooo\fgf n{}\)\,F\(\fgf1{}\ooo\fgf n{}\)\)
\ee
where:
\bel{2.65}
R F\(\fgf1{}\ooo\fgf n{}\)=F\(\fgf1{r}\ooo\fgf n{r}\)
\ee
\i[Tf(4)] for any $\tau \in K_\b$ the random elements $\<\phi,\d_\tau
\tens f\>$ (defined as a unique limits in $L^p(d\mu^\b)$ sense
$\lim_{\epsilon \downarrow 0} \<\phi,\d_\tau^\epsilon \tens f\>$, for any
mollifier $\d_\tau^\e \rightarrow \d_\tau$) exists and moreover the
moments
\bel{2.66}
\mu^\b\(\prod_{i=1}^n e^{i\<\phi,\d_{\tau_i}\tens f_i\>}\) \equiv
G_{f_1\ooo f_n}^{(\mu)}(\tau_1\ooo\tau_n)
\ee
are continuous in $\dt^n \in K_\b\xn\,$; $\dtilde f^n \in \Drd\xn $.
\i[Tf(5)]
a generalized random thermal field $\mu$ is Euclidean invariant iff the
moments $G^{(\mu)}$ are invariant under the natural action of $E(d)$ in
$\Drd$.
\i[Additionally:] a generalized random thermal field $\mu$ will be
called tempered iff the moments \r{2.66} are tempered distributions,
$L^p$-continuous iff the moments \r{2.66} are $L^p$-continuous, etc.
\ed
\end{df}
\begin{prop}\label{p2.9}
\ben
\i Let $(\xi_t)_t$ be tempered thermal process with the temperature $\b$.
There exists a unique (up to the unitary equivalence) $W^*$ KMS structure
$$
\CC^\xi=\(\H^\xi,\Om^\xi;\a_t^\xi;\,\pi^\xi\:{\cal A}(S(\Rd))\to
L(\H^\xi),\m^\xi\)
$$
where
$$
\m^\xi=W^*-\{\a^\xi_{t_1}(\pi^\xi(W(f_1)))\,\ldots
\a^\xi_{t_n}(\pi^\xi(W(f_n)))\}
$$
with $f_i=\o f_i \in S(\Rd)$ real, whose Euclidean Green functions
restricted to $\pi^\xi({\cal A}(S(\Rd)))$ coincide with the moments
$G_{\void\ooo\void}^\xi(\tau_1\ooo\tau_n)\,$, i.e. for any
$-\b/2\leq \tau_1 \leq \ldots \leq \tau_n \leq \b/2$:
\bel{2.67}
\ba{l}
\dsp \<\Om^\xi;\,\apwx{\tau_n}{f_n}{\tau_1}{f_1}\Om^\xi\>\\
\qquad\dsp = G_{f_1\ooo f_n}^\xi(\tau_1\ooo\tau_n)\\
\qquad\dsp= E e^{i\xf11}\,\ldots e^{i\xf n n}
\ea
\ee
\i Let $\mu$ be a tempered thermal field ( at the temperature $\b$).
There exists a unique (up to a unitary equivalence) $W^*$ KMS structure
$$
\CC^{(\mu)}=\(\H^{(\mu)},\Om^{(\mu)};\a_t^{(\mu)};\pi^{(\mu)}\in
\fn{Hom}*( A(S(\Rd)),\,L(\H^{(\mu)}));\,\m^{(\mu)}\)
$$
where
$$
\m^{(\mu)}=W^*-\{\a^{(\mu)}_{t_1}(\pi^{(\mu)}(W(f_1)))\,\ldots\a_{t_n}^{(\mu)}(\pi^{(\mu)}(W(f_n)))\}
$$
$$
t_1\ooo t_n\in \RR,\quad f_1\ooo f_n \in S(\Rd);\quad  f_i=\o f_i \}\,,
$$
whose Euclidean Green function restricted to
$\pi^{(\mu)}\({\cal A}(S(\Rd))\)$ coincide with
$G_{\void\ooo\void}^\mu(\tau_1\ooo\tau_n)\,$.
\i If the tempered random thermal fields $\mu$ is the path space measure of a
tempered process $(\xi_t)_t$ i.e.\ if:
\bel{2.68}
\ba{l} \dsp
E e^{i\xf{\tau_1}1}\,\ldots e^{i\xf{\tau_n}n} %\\
\dsp = \mu\(e^{i \fdf1}\, \ldots e^{i\fdf n} \)
\ea
\ee
for all $\tau_1\ooo\tau_n\in K_\b$; $f_1\ooo f_n\in S(\Rd)$ then the
$W^*$-KMS systems $\CC^{(\xi)}$ and $\CC^{(\mu)}$ coincide.
\een \end{prop}
\proof
Let $(\xi_t)_t$ be a given tempered thermal process at the temperature
$\b>0$. It follows from the definition \ref{d2.7} that the moments
$G_{f_1\ooo f_n}^{(\xi)} (\tau_1\ooo\tau_n)$ define on the abelian
sector ${\cal A}(\SRd)$ of the Weyl algebra $\W(\SRd)$ a system of
functions fulfilling EG(1)--EG(6) with modified EG(1)(ii):
\bd
\i[EG(1)(ii)']: the functionals
$$
G_{\void\ooo\void}^{(\xi)}(\ttt)\: \SRd \xn \ni (\fff) \mapsto
G_\fff^{(\xi)}(\ttt)
$$
are continuous and
$\dsp
\left|G_\fff^{(\xi)}(\ttt)\right| \leq 1
$
\ed
and possible lack of EG(5)(ii). Also EG(3) should be properly modified.
All these modifications however do not affect seriously
the construction presented
in the step 1 of the proposition \ref{p2.5}. Proceeding analogously to
step 1 of proposition \ref{p2.5} we can construct $\CC^\xi$. Similarly
we prove the existence of $\CC^\mu$. The identification of $\CC^\xi$ and
$\CC^\mu$ follows from \ref{2.69} and the uniqueness part of (1) and (2).
\endproof

It follows from the results of \cite{32}, stochastic positivity EG(6)
and the proposition \ref{p2.5} that the thermal structure of the free Bose
Gas can be described fully in terms of the corresponding stochastic
thermal structures
\begin{prop}\label{p2.10}
Let $\Ep$ be given by \r{Ep} and let $0<z$ be such that $\sup_p z \exp
\{-\b \Ep\}<1\,$. Then for any $\b>0\,$:
\ben
\i There exists a unique (up to a stochastic equivalence) Gaussian
thermal process $\xito$ with  values in $\DRd$ such that:
\bel{2.69}
E\<\xi^0_t,\,f\>=0\,;\quad E\(\<\xi^0_t,\,f\>\<\xi^0_{t'},\,g\>\) =
\Sbo(|t-t'|,\, f \tens g)
\ee
The process $\xito$ is Euclidean invariant, ergodic and
$L^2$-continuous.
\i There exists a unique (up to a stochastic equivalence) Gaussian
generalized thermal random field $\mu^\b_0$ such that:
\bel{2.70}
\ba{rcl}
\dsp \mu_0^\b(\<\phi,\,f\>)&=&0\,,\\
\dsp \mu_0^\b(\<\phi,\,\d_\t \tens f\>\<\phi,\,\d_{\t'}f\>)
       &=&\Sbo\(|\t-\t'|,\,f \tens g\)\,.
\ea
\ee
The thermal field $\mu^\b_0$ is Euclidean invariant, ergodic and
$L^2$-continuous.
\i The generalized random field $\mbo$ can be identified with the path
space measure of the process $\xito$ i.e.\ for any bounded, cylindric
function $F$ with base $(\t_1,f_1)\ooo(\t_n,f_n)$
\bel{2.71}
E F\(\xf{\t_1}{1}\ooo\xf{\t_n}{n}\) \equiv
\mbo \( F( \fdf1 \ooo \fdf n)\)
\ee
\i Let $\Vbo$ be a Gaussian measure on $\DRd$ with mean zero and the
covariance given by:
\bel{2.72}
\Vbo\(\<\vf,f\>\<\vf,g\>\) =\Cbo(f \tens g)
\ee
Then the measure $\Vbo$ is the \underline{unique} stationary measure
of the process $\xito$ and $\Vbo$ is equal to the restriction of
$\mbo$ to the $\sigma$-algebra at $\t=0\,$, i.e.\
$\mbo{}_{|\Sigma(0)}=\Vbo\,$, where
$$
\Sigma(0)=\sigma\{\<\phi,\d_0\tens f\>\,;\,f \in f\in \Drd\}\,.
$$
Moreover the measure $\Vbo$ is quasiinvariant under the translations by $\Drd$.
\een
\end{prop}

\remarks
Other well known examples of the generalized thermal processes arise in the
study of two-dimensional models of Euclidean (Quantum) Field
Theory (\cite{23,33}) and also in the context of the Euclidean version of
the Bisognano-Wichman theorem \cite{33,34}. Similar stochastic thermal
structures on the abelian sectors of the corresponding algebras of
observables do appear also in the context of (an)harmonic lattice
crystals \cite{21,26,27} and certain spin systems \cite{24,35}.

The common problem of all these examples is to construct a modular
structure on whole algebra of observables from arising stochastic
thermal structures on the abelian sector. In the case of the Free Bose
Gas the complete solution of this problem is given by proposition
\ref{p2.5}.

{}From the assumption $\sup_p |z \exp\{-\b\Ep\}|<1$ it follows that the
operator $(1- z \exp\{-\b\Ep\})^{-1}$ exists in $\LRd$ and is bounded,
strictly positive and self adjoint. Let $\gh^\b(\Rd)$ be the metric
completion of the space $\Drd$ equipped with the inner product
\bel{prod}
\<f,\,g\> \equiv \int \o{f(x)}(1- z e^{-\b\Ep})^{-1}(x-y)g(y)\,dx\,dy\,.
\ee
{}From the simple estimates:
\bel{2.73}
\|f\|_{\LRd} \leq \|f\|_\b \leq \( \inf_p (1- z e^{-\b\Ep})\)^{-1}
\,\|f\|_{\LRd}
\ee
it follows that $\gh^\b$ is essentially equal to $L_2(\Rd)$. Using the
$\LRd$-continuity of the process $\xot$ and estimates \r{2.73} we can
define a version  $\~\xot$ of  $\xot$ obtained by extension of the
index space $\Drd$ onto the space $\gh^\b$. The new process $\~\xot$
is indexed by $K_\b\x \gh^\b(\Rd)$. For any Borel subset
$I \subset K_\b$ we denote by $\S(I)$ the smallest $\s$-algebra
generated by $\left\{(\~\xot,\,f)\,|\,t\in J,\, f\in
\gh^\b(\Rd)\right\}$. For any $t,s \in K_\b$ we will denote by $[t,s]$
the closed interval from $t$ to $s$ in the counterclockwise direction.
The corresponding conditional expectations with respect to the
$\s$-algebras $\S[t,s]$ ($\S(J)$) will be denoted by $E^0_{[t,s]}$
(resp.\ $E^0_J$).
\begin{prop}\label{p2.11} ~~~\phantom{a} \\
\ben
\i For any allowed form of $\Ep$, $z$ such that $|ze^{-\Ep}|<1$ the
corresponding free thermal process $\~\xot$ has two-sided Markov property
on $K_\b$ in the sense that:
\bel{2.75}
E^0_{[s,r]}\circ E^0_{[r,s]} = E^0_{\{r,s\}}\circ E^0_{[r,s]}
\ee
\i Let $\Xi(J) \equiv \s\{\phi (t,f)\,|\,t\in J,\,f\in\gh^\b\}$
be the corresponding $\s$-algebras in $B({\cal D}'(\Kb\x\Rd))$ and let
$\~E^0(J)$ denote the corresponding conditional expectation values. Then
the free thermal random field $\mbo$ has the following two-sided Markov
property on $\Kb$:
\bel{2.76}
\~E^0_{[s,r]}\circ \~E^0_{[r,s]} = \~E^0_{\{r,s\}}\circ \~E^0_{[r,s]}
\ee
\een
\end{prop}
\proof
It follows easily easily that the operator $h^\mu=h_0+\mu 1$ is a
nonnegative selfadjoint operator in $\gh^\b$ (on the same domain as in
$\LRd$). Moreover the covariance operator $\Gbo(t)$ of the process
$\~\xot$ indexed by $\Kb\x \gh^\b_r$ is given by:
\bel{2.77}
\Gbo(t) = e^{-th^\mu} + e^{-(\b-t)h^\mu}
\ee
Applying theorem 4.1 of \cite{32} we conclude the proof of the first
section. The second
part follows easily by identification of $\phi(t,x)$ with $\xt(x)$ given
in proposition \ref{p2.10} and the density of $\Drd$ in the space
$\gh^\b_r(\Rd)$.
\endproof

\subsection{Local aspects (The case $\Ep=p^2$)}
Let $\Lam\subset\Rd$ be a bounded region with a boundary $\p \Lam$ of a
class at least $C^1$-piecewise. Then, for any $b \in C(\p \Lam)$ the
selfadjoint extension $-\D^b_\Lam$ of the symmetric operator $-\D$ defined
on $C^\infty_0(\Lam)$ can be constructed. It is well known that the arising
semigroup $\{e^{-t\D^b_\Lam}\,,\,t \geq0\}$ is positivity preserving on
$L_2(\Lam)$ therefore there exists a stationary, with independent
increments Markov process $\B^\s_\Lam(t)$ with values in $\o\Lam$ for which
the kernel $K_t^{(\Lam,b)}$ of $e^{-t\D^b_\Lam}$ plays the role of the
transition function.

Let $\W_\Lam$ be the local Weyl algebra built over the space $L_2(\Lam)$ and
let $\W^F_\Lam$ be its Fock space realisation in the Fock-Bose space
$\G_{-1}(L_2(\Lam)$. In particular we have
\bel{2.77a}
W_\Lam^F(f)=e^{i[a^+_\Lam(\o f) + a_\Lam(f)]} = e^{i \vf_\Lam(f)}
\ee
where $a_\Lam$ and $a_\Lam^+$ are standard annihilation and creation
operators in $\G_{-1}(L_2(\Lam)$.

Let $P^{(\Lam,b)}(d\lam)$ be the spectral measure for the operator
$-\D^b_\Lam$. Then, we can define finite volume thermal state
$\om_0^{(\Lam,b)}$ on $\W^F_\Lam$ by the formula
\bel{2.78}
\om_0^\Lb(W_F(f))= \exp {-\half C^\b_{0,\Lb}(f)}
\ee
where
\bel{2.79}
C^\b_{0,\Lb}(f) \equiv \<f|C^\b_{0,\Lb}(f)\>_{L_2(\Lam)}\,;
\ee
\bel{2.80}
\^C^\b_{0,\Lb}(f) = \int P^\Lb (d\lam) { 1+z e^{-\b \lam} \over 1 - z e^{-\b
\lam}}
\ee
It is well known (see i.e.\ \cite{17}) that for any monotonic sequence
$(\Ln)_n$ of bounded regions in $\Rd$ and
with sufficiently regular boundaries $\p \Ln$
 tending to $\Rd$ by inclusion and for any sequence
$b_{\p\Ln}\in C(\p\Ln)$ we have the weak convergence:
$\lim_{n\rightarrow \infty}\^C^\b_{0,(\Ln,b_n)}=C_0^\b$ if $z\in (0,1)$.
The corresponding GNS construction applied to
$(\W_\Lam^F(\void;\,\om_0^\Lb)$
leads again to $W^*$-KMS system
$\CC_0^\Lb=(\Ho^\Lb;\,\pi_0^\Lb;\,\Omo^\Lb;\,\a_t^\Lb;\,\pi_0^\Lb(W^F_\Lam)\2)$
and the corresponding Green functions can again be easily computed and
the analicity properties similar to those of $\GG_0$ established. In
particular the corresponding Euclidean Green functions $\E \GG_0\lbl$
again fulfill the system of axioms $EG(1)--EG(4)$ and $EG(6)$, therefore
the whole discussion from the subsection 2.1 can be repeated with
obvious modifications.
\begin{lem}\label{l2.12}
Let $z=e^{-\b\mu}$ be sufficiently small and let $(\Ln)$ be a monotonic
sequence of bounded convex regions in $\Rd$ with boundaries $\p\Ln$ of
class at least $C^3$ and with mean curvatures uniformly bounded. Then
for any choice of $b_n \in C(\p \Ln)$, any $f_1\ooo f_m \in L_2(\Lam)$,
$\ds^m\in T^\b_n$ we have the convergence
\bel{2.81}
lim_{n \rightarrow \infty} \E G_0(\Ln,b_n)\((s_1,f_1)\ooo(s_m,f_m)\)
= \E G_0^0\((s_1,f_1)\ooo(s_m,f_m)\)
\ee
\end{lem}
\proof
The monotonicity in the boundary conditions:

If $b_1(x) \leq b_2(x)$ for all $x \in \p \Lam$, then
\bel{2.82}
K_t^{(\Lam,b_1)}(x,\,y) \geq K_t^{(\Lam,b_2)}(x,\,y)
\ee
for all $x,y \in\Lam$, $t>0$. Therefore
\bel{2.83}
\sup_{b\in C(\p\Lam)} |\K_t^\Lb(x,\,y)-\K_t(x,\,y)|
= |\K_t^{(\Lam,0)}(x,\,y)-\K_t(x,\,y)|
\ee
for all $t,x,y \in\Lam$, where $\K_t^{(\Lam,0)}$ is the kernel of the
semigroup $\{e^{-t\D_\Lam^N}(x,\,y),\,t \geq 0 \}$, where $\D^N_\Lam$
correspond to the Neumann boundary condition. By the (rough) estimate of
\cite{36} we have with our assumptions on $(\Ln)$:
\bel{2.84}
\left|\K^{(\Ln,b^{\p\Ln})}_t(x,y) - \K_t(x,y)\right|
\leq C e^{\lam t} t^{-d/2}\exp\left\{-c\({ d(x,\Ln^c)^2+d(y,\Ln^c)^2 \over
4t }\)\right\}
\ee
for all $x,y \in \Lam$, $t \in \RR$, where $C$, $c$ and $\lam\geq0$ are
some constants.

It is due to the quasifree nature of the states $\om_0^\Lb$ that it is
enough to consider the one time Green function only
\bel{2.85}
\ba{l}
\dsp \left|\E G^0(\Ln,b_n)((0,f_1),(s_1,f_2))-\E
        G^0((0,f_1),(s_1,f_2)) \right|\\
\dsq \leq \left| \exp\(i\s(f_1,e^{is_1 h^\mu(\Ln,b_n)}f_2)\)\right|\\
\dsqq    \left|\Sbo(s_1\,|\,f_2\tens f_2) -
        e^{i[\s(f_1,e^{is_1h_\Lam^\mu}f_2)-\s(f_1,e^{is_1 h^\mu}f_2)]}
        \Sbo(s_1, f_1\tens f_2)\right|
\ea
\ee
It is well known that $\limni e^{it(-\D_{\Ln}^{b_{\Ln}} +\mu 1)}
= e^{it(-\D+\mu 1)}$ strongly in $\LRd$, therefore we shall omit the
symplectic factor in the last formula, concentrating attention on:
\bel{2.86}
\ba{l}
\dsp \left|^\Lbn\Sbo(s|\,f_1\tens f_2)-\Sbo (s|\,f_1\tens f_2)\right|\\
\dsq \leq \sumno z^{n+s/\b}\int dx\,dy\,f_1(x)f_2(y)
     \left|\K^\Lbn _{(\b n+s)} (x,y) -\K_{(\b n +s)}(x,y)\right| \\
\dsq + \sumno z^{n+1-s/\b}\int dx\,dy\,f_1(x)f_2(y)
     \left|\K^\Lbn _{(\b(n+1)-s)} (x,y) -\K_{(\b(n+1)-s)}(x,y)\right| \\
\ea
\ee
Therefore localizing firstly $f_1$, $f_2$  and
taking into account \r{2.84} we obtain
\bel{2.87}
\limni {}^\Lbn\Sbo(s,\,f_1\tens f_2) =\Sbo(s|\,f_1\tens f_2)
\ee
 provided $e^{-\b\mu}e^\lam<1$.
\endproof

\remarks
The restriction $e^{-\b\mu}e^\lam<1$ is by no doubts only an artifact of
the rough estimate \r{2.84} used. It is natural to expect that actually this
lemma is valid for all $0<z<1$. For a Dirichlet boundary condition the
constant $\lam$ can be taken equal $0$ and this gives the result of the
independence of the limiting Green functions of the Dirichlet boundary
condition in the full noncritical interval $z\in(0,1)$.

The finite volume, conditional thermal processes (resp. thermal random
fields) will be denoted by $\xi_t^{(\Lam,b_{\p\Lam})}$ (resp.
$\mu^{(\Lam,b_{\p\Lam})}_0$).

Having established properties EG(1) $\div$ EG(4) of the corresponding
Euclidean Green functions $\E\GG_0^\lbl$ (resp $^{AE}\GG_0^\lbl$) we can
construct again three different {\sl a priori} $W^*$-KMS structures:
$\E\CC_0^\lbl$, $^{EA}\CC_0^\lbl$, and the basic GNS system
$\CC_0^\lbl$. It appears that all the claims of a properly modified
proposition \ref{p2.5} are still valid and the proof is almost identical
with the exception of the lemma \ref{l2.6} which is replaced by the
lemma \ref{l2.13}.

Let $\Lam$ be a bounded, open and connected region in $\Rd$, $d \geq 2$
with a smooth boundary. Let us define $-\D^b_\Lam(f)=-\D f$ for $f\in
C^2_0(\Lam)$, where ${\cal D}(-\D^b_\Lam)$
consists of those $f\in L^2(\Lam)$ which satisfy
\ben
\i[a)] $f \in C^2(\Lam)\,,$
\i[b)]
$\p^n f(x) = b(x) f(x) $ for $x\in \p \Lam\,,$
%\i[c)] $\varp f{x_i} \in L^(\Om)\,,\quad \D f \in L^(\Om)\,.$
\een
with $\p^n$ being normal inward
derivative.
It follows that $-\D^b_\Lam$ for $b\in C^1(\p \Lam)$ is densely defined,
symmetric and strongly positive.
Let $\^L^b$ be the Friedrichs' extension of $-\D^b_\Lam $
to a self-adjoint
operator. Then at it is well known (see i.e. \cite{37}) the spectrum
of a selfadjoint $L^b$ is
purely discrete and all eigenfunctions of $L^b$ are real valued.
Moreover the semigroup $\dsp e^{-tL^b_\Lam}$ is of trace class.

 It is well known (see i.e.\ \cite{37}) that $\^L^b_\Lam$
 possesses  real-valued
eigenfunctions $\{u_k\}$ associated with eigenvalues $0>\lam_1\geq
\lam_2\geq \,\ldots\,$. Moreover $\{u_k\}_{k=1}^\infty$ form a
complete set in $L^2(\Lam)$.
\begin{lem}\label{l2.13}
A linear space generated by functions $e^{it\^L^b}f$, where $t\in \RR$ and
$f=\o f$, $f\in L^2(\Lam)$ is dense in $L^2(\Lam)$.
\end{lem}
\proof
It is enough to show that for every
$$
f=\sum_{k=1}^n z_k u_k\,,\quad z_k\in\CC
$$
there exist $t_0,t_1\ooo t_m\in\RR$, $f_0=\o f_0$, $f_1=\o
f_1,\,$\ldots,$f_m=\o f_m$ from $\LOm$ such that
$$
f=\sum_{j=0}^m e^{it_j\^L^b}f_j\,.
$$
We exploit the fact that $e^{it\^L^b}=\sum_{k=1}^\infty e^{it\lam_k }P_k$,
where $P_k$ is the one-dimensional projector onto $u_k$.

Let $z_k=a_k+i\, b_k\,$, $a_k$ $b_k\in\RR\,$. Let us define
$t_0=0$, $f_0=\sum_{k=1}^n a_k u_k$, $m=2n$, $t_j=-t_{j+n}$ for $j=1\ooo
n$ and
$$
f_j=\left\{ {\dsp \half {b_j \over \lam_j} u_j \quad \mbox{for } j=1\ooo n
            \atop
            \dsp  -\half {b_{j-n} \over \lam_{j-n}} u_{j-n} \quad
            \mbox{for } j=n+1 \ooo m\,.
            }\right.
$$
Then
$$
\sum_{j=0}^{2n} e^{itj\^L^b} f_j = \sum_{k=1}^n a_k u_k
+ \half \sum_{j=1}^n {b_j\over \lam_j}\(e^{it_j\^L^b}-e^{-it_j\^L^b}\)u_j
$$
but
$$ e^{it_j\^L^b}-e^{-it_j\^L^b} = 2i\sum_{k=1}^\infty (\sin t_j \lam_k)
P_k\,.
$$
So by putting $t_j={\pi \over 2} {1 \over \lam_j} $ we obtain that
$$
\sum_{j=0}^{2n} e^{it_j\^L^b} f_j =
\sum_{k=1}^n a_k u_k + i \sum _{j=1}^n b_j u_j = \sum_{k=1}^n z_k u_k\,.
$$
\endproof
In the sequel we shall need also the following Feynman-Kac formulae:
\begin{prop}\label{p2.14}
Let $\Ep=p^2$, and let $0<z<1$.
\ben
\i For any $f=\o f \in L_2(\Lam)$, $b\in C_+(\p \Lam)$
\bel{2.88}
\ba{l}
\dsp {\Tr_\glam e^{i\vf_\Lam(f)} \G_{-1}\(e^{-\b(\D^b_\Lam+\mu 1_\Lam)}\)
\over
\Tr_\glam\( \G_{-1}\(e^{-\b(\D^b_\Lam+\mu 1_\Lam)}\)\)
}\\
\qquad\dsp\equiv \om_0^\Lb(W_F(f))\\
\qquad\dsp\equiv E e^{i \<\xi_t^\Lb,\,f\>} = \mu_0^\Lb\(e^{i\<\phi,\,\d_0 \tens
f\>}\)
\ea
\ee
\i For any $-\b/2 \leq \t_1 \leq\, \ldots \leq \t_n\leq \b/2$, $f_1=\o
f_1$, \ldots,$f_n=\o f_n \in L_2(\Lam)$
\bel{2.89}
\ba{l}
\dsp{\Tr_\glam \(
\a^\Lb_{i\t_1}(\Epo(W^F(f_1)))\,\ldots\E\a^\Lb_{i\t_n}(\Epo(W^F(f_n)))\)
\G_{-1}\(e^{-\b(\D^b_\Lam+\mu 1_\Lam)}\)
\over
\Tr_\glam\(\G_{-1}(e^{-\b(\D^b_\Lam+\mu1)})\)} \\
\quad \dsp= \EE \(\prod_{i=1}^n e^{i\<\xi^\Lb_{\t_i},\,f_i\>}\)\\
\quad \dsp= \mu^\Lb_0\(\prod_{l=1}^n e^{i\<\phi,\,\d_{\t_i}\tens
f_i\>}\)\,,
\ea
\ee
\i For any sequence $(\Lam_m,\,b_{\p\Lam_m})$ as in lemma \ref{l2.12},
 any $-\b/2 \leq \t_1 \leq\, \ldots \leq \t_n\leq \b/2$,
$f_1$, \ldots,$f_n \in \LRd$ real and sufficiently small $z$ limits
$$
\lim_{m \ra \infty}  \EE \(\prod_{i=1}^n
e^{i\<\xi^{(\Lam_m,\,b_m)}_{\t_i},\,f_i\>}\)
$$
$$
(\mbox{resp. } \lim_{m \ra \infty}
\mu^{(\Lam_m,\,b_{\p\Lam_m})}_0\(\prod_{i=1}^m e^{i\<\phi,\,\d_{\t_i}\tens
f_i\>}\)\,),
$$
exist and are equal to
$$
\EE \(\prod_{l=1}^m e^{i\<\xi^0_{\t_l},\,f_l\>}\)
$$ resp
$$
 \mu^\b_0\(\prod_{l=1}^m e^{i\<\phi,\,\d_{\t_l}\tens f_l\>}\)\,.
$$
\een
\end{prop}

\setcounter{equation}{0}
\setcounter{section}{2}
\section{Gentle perturbations of the free Bose Gas: Thermodynamic limits
on the abelian sector}

We shall study the thermodynamic limits of the multiplicative-like
perturbations of the free thermal field $\mu_0^\bm$ given by the
following perturbations:
\bel{3.1}
\mu_{\Lam,\e}^\bm (d\phi) =Z\1_\Lam \exp W_\Lam(\phi_\e)\mu_0^\bm(d\phi)
\ee
where the interactions $W_\Lam(\phi_\e)$ will be of the following form:
\bd  \i[(LGP)]  the local gentle perturbations:
\bel{3.2}
W_\Lam^L(\phi_\e)=\lam \int d \rho (\a) \int_0^\b d\tau \int_\Lam
\,: e^{i\a \phi _\e(\t,x)}:\, dx
\ee
where:\\
$\dsp :e^{i\a\phi_\e(\t,x)}:\,\,= \exp {\a^2 \over 2} S^\b_\e(0,x) \exp
i\a \phi_\e(\t,x)$\\
$d\rho$ is a complex, bounded measure with a compact support and such that
$\o{d\rho(\a)}=d\rho(-\a)$;\\
$\dsp \phi_\e(\t,x) =(\phi * \chi_\e)(\t,x)$, where $(\chi_\e)_\e>0$ is
a positive mollifier i.e.\ $0\leq \chi_\e \in C_c^\infty (\Rd)$, with
support of size smaller than $\e$ and such that $\int\limits_\lam
\chi_\e(x)\,dx =1$;\\
$\lam$ is the strength of the perturbation
\i[(nLGP)] the nonlocal gentle perturbations
\bel{3.3}
\ba{rcl}
\dsp W_\Lam^{nl}(\phi_\e) &=&\dsp \lam \int_0^\b d\t\int d \rho(\a)\,
d\rho(\a') \\
&&\dsp \intl_\Lam dx \intl_\Lam dy
:e^{i\a\phi_\e(\t,x)}:\,V(x-y):\,\,:e^{i\a\phi_\e(\t,y)}:
\ea
\ee

where $\lam$, $d\rho$, $\phi_\e$ are as in the local case, the kernel
$V$ is chosen to be $L_1$ integrable function.
\ed
\begin{lem}\label{l3.1}
For both choices (LGP) and (nLGP) the thermodynamic stability bound
\bel{3.4}
Z_\Lam=\int d\mu_0^\bm \exp W_\Lam (\phi_\e) \leq \exp C\cdot\left|\Lam\right|
\ee
holds, where $C$ is some constant depending on the details of the
perturbations.
\end{lem}

\proof
We shall consider only the case (nLGP). By simple Gaussian calculations
we obtain:
\bel{3.5}
\ba{l}
\dsp \int d\mu_0^\bm(\phi)W_\Lam^{nl}(\phi_\e)^n \\
\dsp \quad = \lam ^n\int _0^\b d\t|_1^n\int d\rho(\a)_1^n \int
d\rho(\a')_1^n\intl _\Lam d x
|_1^n \intl_\Lam dy |_1^n\\
\dsp \qquad \prod _{i=1}^n V(x_i-y_i)
\exp {-\half \sum _{{i,j=1\atop i\neq j}}^n\a_i\a_j S^\b_\e
(\t_i-\t_j,x_i-x_j)}
\\
\dsp \qquad \exp {-\half \sum _{{i,j=1\atop i\neq j}}^n \a'_i \a'_jS^\b_\e
(\t_i-\t_j,y_i-y_j)}
\exp {-\half \sum _{i,j=1}^n \a_i\a'_j S^\b_\e (\t_i-\t_j,x_i-y_j)}
\ea
\ee
Using the positive definitness of $S^\b_\e$ we can estimate:
\bel{3.6}
\ba{l}
\dsp \left| \int d\mu^\bm_0(\phi)W^{nl}_\lam(\phi_\e)^n\right|
\dsp\quad =|\lam|^n\b^n (\fn{Var} \rho)^{2n} \exp \(2n S^\b_\e(0)\)
\|V\|_1^n \,|\Lam|^n
\ea
\ee
which shows the bound \r{3.4} with
\bel{3.7}
C=|\lam|\,\b(\fn{Var} \rho)^{2} \exp \(2 S^\b_\e(0)\)
\|V\|_1\,.
\ee
Moreover it follows that $Z_\Lam$ are entire functions of the coupling
constant $\lam\in\CC$.
\endproof

The characteristic functionals of the perturbed measures $\mu_{\Lam,\e}^\bm$
can be written in the following forms
\bd
\i[(LGP) case]
\bel{3.9}
\ba{rcl}
\dsp \mu_{\Lam,\e} ^\bm\(e^{i \<\phi, g \tens f\>}\) &=&\dsp \exp -\half \Sbo
(g \tens f
| g \tens f) \sum _{n \geq 0} {1 \over n!}  \int _0 ^\b d\t|_1^n \int
d\rho(\a)|_1^n \\
&&\dsp \cdot \intl_\Lam dx |_1^n  \prod _{i=1}^n \[ e^{-i\a_i(g\tens f) * \Sbe
(\t_i,x_i)} -1\]
\rho_{\Lam,\e}(\t,x)_1^n
\ea
\ee
where:
\bel{3.10}
\rho_{\Lam,\e}(\t,\a,x)|_1^n=\lam ^n \int d \mu_\Lam^\bm (\phi)
\prod_{l=1}^n\,:e^{i\a_l\phi_\e(\t_l,x_l)}:\,,
\ee
\i[(nLGP) case]
\bel{3.11}
\ba{l}
\dsp \mu_{\Lam,\e} ^\bm\(e^{i \<\phi, g \tens f\>}\) = \exp -\half \Sbo (g
\tens f
| g \tens f)
 \sum _{n \geq 0} {1 \over n!} \int _0 ^\b d\t|_1^n \int d\rho(\a)|_1^n \\
\qquad \dsp \int d y|_1^n \int d\rho(\a'_i)|_1^n \int d (y_i)_1^n
 \prod _{l=1}^n V(x_i-y_i) \\
\qquad \dsp \prod _{i=1}^n \[ e^{-i\a_i(g \tens f) * \Sbe (\t_i,x_i)
-\b_i(g \tens f)*\Sbe} -1\]\s_{\Lam,\e}\(\t,(\a, x)_1^n,\,(\b, y)_1^n\)
\ea
\ee
where:
\bel{3.11a}
\s_{\Lam,\e}\(\t_1^n,(\a, x)_1^n,\,(\b, y)_1^n\)=\lam ^n \mu_\Lam^\bm
\(\prod_{l=1}^n\,:e^{i\a_l\phi_\e(\t_l,x_l)}:\,
\prod_{l=1}^n\,:e^{i\b_l\phi_\e(\t_l,y_l)}:\,\)
\ee
\ed
Employing the integration by parts formula we obtain the following
equalities
\bd
\i[(LGP)]
\bel{3.11b}
\ba{l}
\rho_{\Lam,\e}(\t,x)_1^n =\dsp \lam^n \exp{ -\sum_{i=2}^n \Sbe (\t_1-\t_i |
x_1 -x_i )\a_i \a_1} \\
\qquad \dsp \mu_{\Lam,\e}^\bm \(\prod_{l=2}^n \,: e^{i\a_i\phi_\e(\t_l,x_l)}:\,
\exp \left\{\lam\int_0^\b d \t \intl_\Lam dx \int
d\rho(\a)\right.\right. \\
\qquad \dsp \left.\left.\[e^{-\a \a _1\Sbe(\t_1-\t;x_1-x)} -1\]
 \,: e^{i\a \phi_\e(\t,x)}:\,\right\}\)
\ea
\ee
\i[(nLGP)]
\bel{3.11c}
\ba{l}
\dsp \s_{\Lam,\e}\((\t,\a,x)_1^n, (\t,\b,y)_1^n\)\\
\quad=\dsp \lam^n \exp {-\sum_{i=2}^n \a_1 \a_i\Sbe(\t_1-\t_i|x_1-x_i)}
        \exp{-\sum_{i=2}^n \b_1 \b_i\Sbe(\s_1-\s_i|y_1-y_i)} \\
\qquad \dsp \mu_{\Lam,\e} ^\bm \(\prod_{l=2}^n \,:e^{i\a_l\phi_\e(\t_l,x_l)}:\,
                \prod_{l=2}^n \,:e^{i\b_l\phi_\e(\s_l,y_l)}: \,\right. \\
\qquad \dsp \exp \left\{\lam \int_0^\b d\t \int
d\lam(\a)\,d\lam(\b)\intl_\Lam dx\intl_\Lam dy \right.\\
\qquad \dsp \[e^{-\a\a_1\Sbe(\t_1-\t,x_1-x)}
e^{-\b\b_1\Sbe(\t_1-\t,y_1-y)}-1\]\\
\qquad \dsp\left.\left. \,:e^{i \a \phi_\e(\t,x)}:\,V(x-y)
\,:e^{i\b\phi_\e(\t,y)}:\,\right\}\)
\ea
\ee
\ed
in which after an convergent expansion in powers of $\lam$ we recognize
the well known \cite{22} Kirkwood-Salsburg-like equalities that hold
between the correlation functions. A straightforward application of the
contraction map principle \cite{22} or the methods of the dual pairs of
Banach spaces \cite{38} leads to the proof of the following
proposition in the (LGP) case.
\begin{prop}\label{p3.1} {\bf (LGP)} \ben
\i For $|\lam| < \lam_0 (LGP)$, where
\bel{3.11d}
\lam_0(LGP) = \exp\(-\a_*^2S_\e(0,0)-1\) C_\e^{l-1}\,,
\ee
where
$$C_\e^l\equiv \sup_{\a'} \int_0^\b d\t \int d|\lam|(\a) \int dx
|e^{\a\a'S_\e(\t,x)}-1|\,,
$$
$$
\a_*^2=\sup \{\a^2 \in \supp d\lam \}\,,
$$
the unique thermodynamic limits
$$ \lim_{\Lam \uparrow \Rd } \rho_{\Lam,\e} (\t,\a,x)_1^n \equiv
\rho_\e(\t,\a,x)_1^n
$$
exist in the sense of locally uniform convergence. The limiting
correlation functions $\rho_\e(\t,\a,x)_1^n$ are continuous, translationally
invariant and have cluster decomposition property. Moreover, they are
analitic functions in $\lam$ for $|\lam| <\lam_0(LGP)$,
\i Let
$$
\lam \in \{z|z\1 \not\in  \s_\xi(\K) \} \cap \{|z|<\xi\} $$
where $K$ is the corresponding infinite-volume KS-operator,\\
$\s_\xi(K)$ is the spectrum of $K$ in the corresponding Banach space
$B_\xi$ (compare with \cite{38,39}).

Then for any such $\lam$ the unique thermodynamic limits
$$
\rho_\e(\t,\a,x)_1^n =\lim_{\Lam \uparrow R^D} \rho_{\Lam,\e} (\t,\a,x)_1^n
$$
exist in the sense of locally uniform convergence and are analitic
functions in $\lam$.
\een
\end{prop}
As a simple corollary we obtain:
\begin{prop}\label{p3.2}{\bf (LGP)}
\ben
\i For $\lam \in \CC$ as described in point 1. or 2. of Prop.\
\ref{p3.1} the weak limit $d\mu^\lam_\e$ of the measure
$d\mu^\b_{\Lam,\e}$ exists and the limiting measure $d\mu^\lam_\e$ is
periodic in $\b$, symmetric on $K_\b$, OS-positive on $K_\b$.
Moreover, $d\mu^\lam_\e$ is (weakly) analitic in $\lam$ perturbation of
the free measure $d\mu_0^\b$.
\i For $|\lam|< \lam_0(LGP)$ the limiting measure $d\mu^\lam_\e$ is
translationally invariant with respect to the translations of $\Rd$ and
is ergodic under the action of this group.
\i For $\lam$ as in 1. the characteristic functional of $d\mu^\lam_\e$ is
given by the following formula
\bel{3.15n}
\ba{rcl}
\dsp \mu^\lam_\e \( e^{i(\phi, g\tens f)}\) &=&\dsp \exp -\half \Sbo (g\tens
f|g\tens
f) \\
&&\dsp \sum_{n\geq 0} {1\over n!} \int d\rho (\a)\, d\t\,dx|_1^n
\prod_{l=1}^n \[ e^{-\a_l\Sbe * (g \tens f) (\t_l, x_l)}-1\] \\
&&\dsp \rho_\e(\t,\a,x)_1^n
\ea
\ee
\een
\end{prop}

A minor modification of the original analysis of the Kirkwood-Salsburg
identities  enables us to control also
thermodynamic limits for nonlocal gentle perturbation \r{3.3} also.
\begin{prop}\label{p3.3}
{\bf (nLGP)} Let $W=(nGLP)$.
\ben
\i For $\lam \in \CC\: |\lam|<\lam_0(nLGP)$ where
$$\lam_0(nLGP)\equiv \exp \( -2 \a^2_* S_\e (0,0) -1\)
\(C_\e^{nl}\)^{-1}\,,
$$
$$
\ba{rcl}
\dsp C_\e^{nl} &\equiv&\dsp \sup_{\g,\g'} \int_0^\b d\t \int d|\lam(\a)\int
 d|\lam|(\a')\\
&&\dsp \int dx\int dy V(x-y) \left| e^{-\a \g S_\e (\t,x)}
e^{-\a'\g'S_\e(\t,y)}-1\right|\,,
\ea
$$
$$
 \a_* =\sup \{ \a \in \supp d\lam\}\,
$$
 $\lim d\mu_{\Lam,\e}^{\b,\mu} = d\mu ^\lam_\e$ exists and the
limiting measure $d\mu^\lam_\e$ is: periodic in $\b$, symmetric on
$\Kb$,
OS-positive on $\Kb$. Moreover, $d\mu^\lam_\e$ is (weakly) analitic in $\lam$
perturbation of
the free measure. The measure $d\mu^\lam_\e$ is $E(d)$ invariant and ergodic
under the translations by $\Rd$. The characteristic functional of
$d\mu^\lam_\e$ is given by the following formula
\bel{3.16n}
\ba{l}
\dsp \mu^\lam_\e\(e^{i(\phi,\,g\tens f)}\) =\exp -\half\Cbo (\gtf|\gtf) \\
\dsqq \sum_{n\geq 0} {1 \over n!} \int
d(\t,x,\a)_1^n\,d(\t',x',\a')_1^n
\prod_{i=1}^n V(x'_i-x_i) \\
\dsqq \prod_{i=1}^n \( \exp\(-\a_i\Sbe * (\gtf)(\t_i,x_i)\)
 \exp\(-\a'_i\Sbe * (\gtf)(\t'_i,x'_i)\) -1\) \\
 \dsqq \s^\lam_\e \((\t,x,\a)_1^n;(\t',x',\a')_1^n\)
 \ea
 \ee
 where:
\bel{3.17n}
\ba{l}
\dsp \s^\lam_\e \((\t,x,\a)_1^n;(\t',x',\a')_1^n\)
\dsq \equiv \lim_{\Lam \uparrow \Rd} \mu_{\Lam,\e}^\lam \( \prod\, :e^{i\a_i
\phi_\e(\t_i,x_i)}:\, \prod _{i=1}^n \, :e^{i\a'_i
\phi_\e(\t'_i,x'_i)}:\)\\
\dsqq = \mu^\lam_\e \( \prod_{i=1}^n\, :e^{i\a_i
\phi_\e(\t_i,x_i)}:\, \prod _{i=1}^n \, :e^{i\a'_i
\phi_\e(\t'_i,x'_i)}:\)\,.
\ea
\ee
\een
\end{prop}

In particular we have obtained the following functional integral
representation of the corresponding multi-time Euclidean Green functions
corresponding to the infinite-volume limit perturbations
 of the free Bose gas in the noncritical regime.
\begin{th}\label{t3.5}
Let: $V_\Lam=(LGP)$ or $V_\Lam=(nLGP)$ and $\lam \in \CC$ be restricted
as in 1. of Prop. \ref{p3.3} or 2. of Prop. \ref{3.2} in (LGP) case. Then the
Euclidean multitime Green functions on ${\cal A}(\gh)$ are given by the
following functional integrals
\bel{3.18n}
\ba{l}
\dsp \E G_{f_1\ooo f_n}^\lam (s_1\ooo s_n) \equiv \lim_{\Lam \uparrow \infty}
\E G_{f_1\ooo f_n}^{\lam,\Lam} (s_1\ooo s_n)
= \intl_{{\cal D}'(\Kb\x\Rd)} d\mu^\lam_\e(\phi) \prod _{i=1}^n
e^{i\<\phi;\,\d_{s_i} \tens f_i\>} \\
\dsq \stackrel{(LGP)}{=} \E G_{f_1\ooo f_n}^0 (s_1\ooo s_n) \\
\dsqq \sum_{n \geq 0}^\infty {1\over n!} \int d(\t,x,\a)_1^n \prod_{i=1}^n
\left\{ \exp -\a_i\Sbe*(\sum_{l=1}^n \d_{\o s_l}\tens f_i)(x_i) -1
\right\} \\
\dsqq \rho_\e((\t,x,\a)_1^n) \\
\dsq \stackrel{(nLGP)}{=} \E G_{f_1\ooo f_n}^0 (s_1\ooo\s_n) \\
\dsqq  \sum_{n \geq 0}^\infty \int d(\t,x,\a)_1^n \, d(\t',x',\a)_1^n
 \prod_{i=1}^n V(x_i-x'_i) \\
\dsqq \exp -\Sbe *(\sum_{l=1}^n \d_{\o s_l}\tens f_i)(x)
         -\a'\Sbe*(\sum_{l=1}^n \d_{\o s_l}\tens f_i)(x') \\
\dsqq \mu_{\Lam,\e}^\lam \( \prod_{l=1}^n  \,:\exp i\a_l\phi_\e
(\t_l,x_l) : \,\cdot \,:\exp i\a'_l \phi_\e(\t'_l,x'_l):\)
\ea
\ee
\end{th}

Some elementary albeit fundamental for the purposes of the present paper
properties of the system are collected in the following proposition:
\begin{prop}\label{p3.6}
Let $\{\E G_{f_1\ooo f_n}^\lam (s_1\ooo\s_n)\}$ be a collection of the
Euclidean-multitime infinite volume Green functions constructed in
Theorem \ref{t3.5}. Then they can be extended by
continuity to the Abelian sector $\Ah$ of the Weyl algebra $\Wh$ and the
continued Green functions denoted by the same symbol obey
properties EG(1)$\div$EG(5)i.
\end{prop}
\begin{cor}\label{c3.7}
Let $|\lam| <\lam_0(LGP)$ (for the case (LPG)) and $|\lam| <\lam_0(nLGP)$
(for the case (nLGP)). Then the following perturbation expansions are
convergent:
\bel{3.19g}
\ba{rcl}
\dsp \E G_{f_1\ooo f_n}^\lam (s_1\ooo s_n)\}
&\stackrel{(LGP)}=&\dsp \sum_{n \geq0} {\lam^n \over n!}
\int\cdots\int\limits_{\Kb\x\Rd\x\RR}
d\t\,dx\,d\rho(\a)|_1^n \\
&& \dsp \cdot \< \prod _{l=1}^n  e^{\<\phi,\d_{s_l} \tens f_i\>}
;\, :e^{i\a_1\phi_\e(\t_1,x_1)}:\,\,;\void ;\void \>_0^{\b,T}
\ea
\ee
where $\<\void;\void;\void;\void\>_0^{\b,T} $ denote the truncated
expectation values with respect to the free Gas measure $d\mu_0^\b$.
\end{cor}
For a class of gentle perturbations of the free Bose gas stochastic
structure another variety of the existence results can be established
using the methods of \cite{5,40}. For this let us
assume now that our perturbations are of the following forms
\bd
\i[(LGP)$_e$] $ \qquad W_\Lam (\phi_\e) = (3.2)$
\ed
but now $d\rho$ is an even bounded real measure, $\lam \geq 0$ or
\bd
\i[(nLGP)$_e$] $\qquad W_\Lam(\phi_\e)=(3.3)$
\ed
where $d\rho$ is also an even bounded real measure and $V\in L_1(\Rd)$ is
assumed to be pointwise nonnegative i.e.\ $V(x)\geq 0$ and $\lam \geq 0$.
\begin{prop}\label{p3.8}
Let $d\mu^\lam_{\Lam,\e}$ be a locally perturbed free Bose Gas measure by
(LGP)$_e$ or (nLGP)$_e$ and let $\lam>0$. Then the following correlation
inequalities of the Fr\"ohlich-Park type are valid.
\begin{eqnarray}
1.&&Z_{\Lam_1\cup\Lam_2} \geq Z_{\Lam_1}\cdot Z_{\Lam_2}\,,\label{c1}\\
2.&&\<\phi^2(\gtf);\,: \cos \a \phi_\e:\,(\t,x)\>^{\lam,T}_{\Lam,\e}
\leq 0\,,\label{c2}\\
3.&&\<e^{t\phi(\gtf)};\,\prod_{i=1}^n: \cos \a_i
\phi_\e:\,(\t_i,x_i)\>^{\lam,T}_{\Lam,\e} \leq 0\,,\label{c3}\\
4.&&\<e^{i\phi(\gtf)};\,\prod_{i=1}^n: \cos \a_i
\phi_\e:\,(\t_i,x_i)\>^{\lam,T}_{\Lam,\e} \geq 0\,,\label{c4}\\
5.&&\<\prod_i \cos \a_i \phi_\e(s_i,x_i) \, \prod_j \cos \b_j
\phi_\e(t_j,y_j)\>^{\lam,T}_{\Lam,\e} \label{c5}
\geq 0\,.
\end{eqnarray}
\end{prop}
\proof
Basically the same as in \cite{5} employing duplicate variable
trick and the elementary trigonometric identities.
\endproof
\begin{th}\label{t3.9}
Let us consider perturbation $(LGP)_e$ or $(nLGP)_e$ of the free Bose
Gas thermal field $d\mu^\b_0$.
\ben
\i
For any $\lam\geq 0$ the unique thermodynamic limit
\bel{3.25g}
\ba{l}
\dsp \lim_{\Lam \uparrow \Rd} \mu^{\lam}_{\Lam,\e} \(\prod_{i=1}^n
e^{i\<\phi,\,\d_{s_i} \tens f_i\>}\)
 \equiv \mu_\e^\lam \( \prod_{l=1}^n e^{i \<\phi,\delta_{s_i}\tens
f_i\>}\) \\
\dsp \equiv \E G^\lam_{f_1\ldots f_n}(s_1\ooo s_n)
\quad \mbox{ for }-\b/2 \leq s_1 \leq \ldots \leq s_n \leq \b/2\,\\
\ea
\ee
exists and the limiting Green functions obey all the properties
EG(1)$\div$EG(5)(i).
\i In particular the following estimates hold:
\ben
\i
\bel{3.26g}
\ba{r}
\dsp \left| S^2_\lam(\ftg | \ftg )  \equiv
\dsp  {d^2 \over i^2
d\a_1\,d\a_2} \E G^\lam_{\a_1 g,\a_2 g}(f,f)_{\left|{\a_1=0 \atop \a_2=0}
\right.}\right| \\
\leq \dsp  \Sbo (\ftg |\ftg)
\ea
\ee
\i
\bel{3.27g}
\ba{r} \dsp
\left| \mu^\lam_\e \(\exp S \int_0^\b d\t\,f(\t) \int dx\, g(x)\phi(\t,x)\)
\right|\\
\dsq \leq \exp \fn{Re} {S^2\over 2} \Sbo(\ftg|\ftg)
\ea
\ee
\i
\bel{3.28g}
\ba{r}
\dsp \left|S^{n,\b}_\lam (f_1\tens g_2\ooo f_n\tens g_n)
\equiv  \mu^{\lam}_\e \(\prod_{i=1}^n\<\phi,f_i\tens
g_i\>\)\right| \\
\dsq \leq \s(n!)^{\half}\prod_{i=1}^n \left| S_0^\b (f_i\tens g_i
| f_i \tens g_i)\right|
\ea
\ee
\een
\een
\end{th}
\proof
{}From the correlation inequality \r{c4} it follows that
$\mu^\lam_\Lam\(e^{i\phi(\ftg)}\)$ monotonously increase in the volume
and that for real $t$ $\mu^\lam_\Lam\(e^{t\phi(\ftg)}\)$ decrease as
$\Lam \uparrow \Rd$. This leads to the statement that the unique limit
$\lim_\Lam \mu_\Lam \(e^{\zeta(\phi, \ftg)}\)$
$\equiv \mu_\infty^\lam\(e^{\zeta(\phi, \ftg)}\)$ exists and obeys the
estimate \r{3.27g}. Then the estimates \r{3.28g} follow by the application of
the
Cauchy integral formula and the analicity in $\zeta$ of
$\mu_\infty^\lam\(e^{\zeta(\phi, \ftg)}\)$. Although the estimate
\r{3.26g}
follows from \r{3.28g} its independent proof follows easily from the
correlation inequality \r{c2} which says that
$\mu^\lam_\Lam(\phi,\ftg)^2$ is monotonously decreasing in the volume.

Integrating by parts on the functional space ${\cal D}'(\Kb\x\Rd)$ with
respect to the measure $d\mu^\lam_\Lam(\phi)$ the following formulae are
obtained:
\bel{3.29g}
\ba{rcl}
\dsp \E G^\lam_{\Lam,\,f_1\ooo f_n}(s_1\ooo s_n)&
\mathop{=}\limits_{GLP_e}&
\dsp \E G^0_{f_1\ooo f_n}(s_1\ooo s_n) \\
&&\dsp \sum_{k \geq 0}^\infty {\lam^k \over k!} \int_{\Lam\x\Kb\x\RR}
d\t\,dx\,d\lam(\a)|_1^k \\
&& \dsp \prod_{i=1}^k \[e^{-\sum_{j=1}^n \a_j\Sbe*(\d_{s_j}\tens f_j)}
-1\] \\
&& \dsp
\mu^\lam_\Lam\( \prod_{j=1}^n :e^{i \a_j \phi_\e(\t_j,x_j)}:\)
\ea
\ee
{}From the correlation inequality \r{c5} if follows that
\bel{3.30g}
\mlL\(:\prod_{i=1}^n \cos\a_i\phi_\e(\t_i,x_i):\)\equiv
C^\lam_\Lam(\a_1,\t_i, x_i|_1^n)
\ee
 monotonously increase in the volume $\Lam$ and because they are
 uniformly bounded
\bel{3.31g}
\left|C_\Lam^\lam(\t_i,x_i|_1^n)\right| \leq \exp \half \b^2 n
\Cbe(0)\,,
\ee
the unique thermodynamic limits $\lim_\Lam \ClL \equiv C^\lam $ exist
pointwise on $(\Kb\x\Rd)^{\tens n}$. From this, the existence of
pointwise limits
\bel{3.32g}
\lim_\Lam \mu^\lam_\Lam \( \prod _{j=1}^n :e^{i\a_j \phi_\e
(\t_1,x_j)}:\) = \mu ^\lam_\e \(\prod_{j=1}^n :e^{i\a_j \phi_e(\t_j,x_j)}:\)
\ee
follows in the same way as demonstrated in \cite{40} by the
application of another correlation inequitity (originally due to Pfister
\cite{41}) not listed in proposition \ref{p3.8} but formulated in
\cite{40} in a similar context. Finally the proven
pointwise convergence is sharpened to the local uniform one by a
standard application of the Mayer-Montrolle identities, see i.e.
\cite{38}. From the obtained convergence the following expression for
the infinite-volume Euclidean Green functions $\E G^\lam_{f_1\ooo f_n}
(s_1\ooo s_n)$ follows easily from \r{3.29g}:
\bel{3.33g}
\ba{rcl}
\dsp \E G^\lam_{f_1\ooo f_n}(s_1\ooo s_n)& =&\E G^0_{f_1\ooo
f_n}(s_1\ooo s_n) \\
&&\dsp \sum_{k \geq 0} {1 \over k!} \int_{\RR_\b\x\Rd\x\RR}
d\t\,dx\,d\lam(\a)|_1^k \\
&& \dsp \prod_{i=1}^k \[e^{-\sum_{j=1}^n \a_i\Sbe*(\d_{s_j}\tens f_j)}
-1\] \\
&& \dsp
\mu^\lam_\infty\( \prod_{i=1}^n :e^{i \a_i \phi_\e(\t_i,x_i)}:\)
\ea
\ee
The case of nLGP$_e$ is analised in a similar way.
\endproof
\remarks
The existence and uniqueness of the thermodynamic limits for the
Euclidean Green functions $\E G^\lam_{f_1\ooo f_n}(s_1\ooo s_n)$
follows easily from the correlation inequality \r{c4} and the uniform
bound:
\bel{3.34g}
\left|\E G^\lam_{f_1\ooo f_n}(s_1\ooo s_n)\right|\leq 1\,.
\ee
%The following corrolaries follow by an applications of the methods
%developed in \cite{} (see also paragraph 2).

%\begin{prop}\label{p3.10}
%a
%\end{prop}

Using the methods based on the analysis of the corresponding
Kirkwood-Salsburg identities one can study the gentle perturbations of
the local, free, conditioned thermal fields described in the section
2.2.

For this goal let us consider a perturbation of the free, conditioned
(by $b_{\p\Lam} \in C(\p\Lam)$), thermal field $\mu_0^{(\Lam,\,
b_{\p\Lam})}$ of the form:
\bel{3.35}
\~\mu^{(\lam,\ b_{\p\Lam})}_{\Lam,\e} (d\Phi)= \~Z^{-1}_{\Lam,\e}
(b_{\p\Lam}) \exp W_{\Lam}(\Phi_\e)\ d\mu_0^{(\Lam,\ b_{\p\Lam})}(\Phi)
\ee
where:
\bel{3.36}
\~Z_{\Lam,\e}(b_{\p\Lam})=\mu_0^{(\Lam,\ b_{\p\Lam})}\(\exp W_{\Lam}\)
\ee
and $W_{\Lam}(\Phi_\e)$ is given by \r{3.2} or \r{3.3}.
\begin{th}\label{t3.10}
Let $(\Lam_\a)$ be any arbitrary net of bounded subsets of $\Rd$ with
the boundaries of class at least $C^3$-piecewise. Additionally we shall
require that the mean curvature of $\p\Lam_{\a}$ is uniformly bounded in
$\a$. Let $(b_{\p\Lam_{\a}}^\a)$ be a sequence of continuous boundary
conditions.

Then for $|\lam| < \lam_0 (LGP)$, if $W_{\Lam} = LGP$ (respectively $|\lam|
 < \lam_0 (nLGP)$, if $W_{\Lam} = nLGP$) the unique thermodynamic limits
 $$
 \lim_\a\~{\mu}_{\Lam,\e}^{(\lam,\ b_{\p\Lam_\a})} \equiv
 \~\mu_\e^\lam
 $$
 exists in the sense of weak convergence and moreover $\~\mu_\e^\lam =
 \mu_\e^\lam$.
 \end{th}
\proof
The method of the dual pair of Banach spaces as explained in \cite{38}
and applied in the similar situation in \cite{39,40} is applied.
\endproof
\par {\bf Remark}\\

The method of \cite{38,39,40} gives the existence and independence on
the classical boundary conditions of the limiting thermal field
$\~\mu_\e^\lam$ in a larger set of $\lam$ (see also point 2 in Prop
\ref{p3.2} above).

As a corollary we have the following result:
\begin{cor}\label{c3.11}
Let $(\Lam_\a)_\a,\ (b_{\p\Lam_\a})_\a$ be as in Theorem \ref{t3.10} and let
${}^A \bbG_\lam (\Lam_\a$, $b_{\p\Lam_\a})$ be the system of the Euclidean
Green
functions corresponding to the gentle perturbations of the local,
conditioned, free $W^*$-KMS structure restricted to the Abelian
sector ${\cal A}(\gh_\Lam)$ of ${\cal W}(\gh_\Lam)$. Then for $\lam$ as
in Theorem
\ref{t3.10} and $0<z<1$ sufficiently small the unique thermodynamic limits of
the corresponding Euclidean Green functions exist and are equal to those
obtained in Theorem \ref{t3.5} and Theorem \ref{t3.9}.
\end{cor}

All the constructed in this section systems of limiting Euclidean Green
functions obey properties $EG(1) \div EG(5)(i)$ and corresponds to some
generalized thermal processes.

Therefore the general reconstruction procedures of \cite{20} applies,
(see Prop\ref{p2.9}), leading to certain $W^*$-KMS structures Further
analysis of the derived $W^*$-KMS structures is contained in the
forthcoming papers.

\sec
\section{Concluding Remarks}
\subsection{}
For the finite volume perturbations of the free thermal field
$\mu_0^{(\b,\mu)}$ the corresponding nonhomogeneous process
$(\xi^{(\lam,\Lam)}_t)_{t\e K_\b}$ has two-sided Markov property on
$K_\b$ in the sense of Prop \ref{p2.11}. The interesting and important question
is whether the homogenous limits $\Lam \uparrow \Rd$ preserve the above
Markov property. For a gentle perturbations of a class of lattice
anharmonic crystals some results on the preservation of the two-sided
Markov property in the thermodynamic limit have been established in
\cite{42}. A constructive route for the verification of the two-sided
Markov property will be formulated below.
\subsection{}
The notion of $DLR$ equations for the gentle perturbations of the
abelian sector of the free Bose Gas in the Euclidean region can be
introduced. For this goal, let us denote by $\Pi(\Lam^C)$ the orthogonal
projector (in the space $\H_0^\b \equiv m.c.\(C(K_\b) \times D(\Rd);
S_0^\b\)$) onto the subspace $\H_0^\b (\Lam^C) \equiv
m.c.\(C(K_\b)\right.$
$\times$ $C_C^{\infty}(\Lam^C);$ $\left.S_0^\b\),$
for $\Lam \subset \Rd$ open and
bounded.

The free thermal kernel $S_0^\b$ is then decomposed as:
\bel{4.1}
S_0^\b = {}^{\Lam^C} S_0^\b + {}^{\Lam^C}\Pi_0^\b
\ee
where:
\bel{4.2}
 {}^{\Lam^C}S_0^\b \equiv S_0^\b \circ \(1 - \Pi(\Lam^C)\);
 {}^{\Lam^C}\Pi_0^\b = S_0^\b \circ \Pi(\Lam^C)
 \ee
 Let $\mu_0^{\Lam^C}$ be a Gaussian random field with the covariance
 given by ${}^{\Lam^C}S_0^\b$. It is clear that the symmetricity and
 $OS$
 positivity on $K_\b$ of the free conditioned Gaussian random field
 $\mu_0^{\Lam^C}$ is preserved and moreover $\mu_0^{\Lam^C}
 \longrightarrow \mu_0^\b$ weakly as $\Lam \uparrow \Rd$.

Let $\S^0(\Lam^C)$ be the ($\mu^0$-complete) $\s$-algebra generated
by the random elements of the form $\<\Phi,\ f\>$, where $f \in \H_0^\b
(\Lam^C)$. Then the conditional expectation values of the measure
$\mu_0^\b$ with respect to the $\s$-algebras $\S^0(\Lam^C)$ are given
by:
\bel{4.3}
E_{\mu_0} \{F| \S^0(\Lam^C)\}(\Psi) = \mu_0^{\Lam^C}\(F(\void +
\Pi_{\Lam^C}^*(\Psi)\)
\ee
for: $\mu_0$ --- a.e. $\Psi \in {{\cal D}'}(K_\b \times \Rd)$, where
\bel{4.4}
\<\Pi^*_{\Lam^C} (\Psi),\ f>\,\, \equiv \,\,<\Psi,\ \Pi_{\Lam^C}(f)\>.
\ee
The corresponding conditional expectation values of the perturbed
measure are:
\bel{4.5}
\ba{l}
\dsp E_{\mu_{\Lam,\e}} \{F | \S^0_{\Lam^C})\}(\Psi)\\
\quad\dsp = {{\mu_0^{\Lam^C}\(F(\cdot + \Pi^*_{\Lam^C}(\Psi)) \exp
W_{\Lam} (\cdot + \Pi_{\Lam^C}^*(\Psi)\)}
\over
{\mu_0^{\Lam^C}\(\exp W_{\Lam}(\cdot + \Pi^*_{\Lam^C}(\Psi)\)}
}\\
\ea
\ee
for $\mu_0$ --- a.e. $\Psi \in {\cal D}'(K_\b \times \Rd)$.

In analogy to \cite{27} (see also \cite{43,44}) we define a classical
thermal Gibbs measure corresponding to the gentle perturbation of the
free Bose Gas as any probabilistic, cylindric Borel measure $\mu$ on
${\cal D}'(K_\b \times \Rd)$ such that
$$
 \mu \circ E_{\mu_{\Lam,\e}}\left\{ \S (\Lam^C)\right\} = \mu
\,\,\,\,\,\,\,\,\,\,\,\,\,\,\,\,\,\,\,\,\,(DLR).
$$
for any open bounded $\Lam \subset \Rd$.

It is evident that any solution of (DLR) defines a thermal random field
in the sense of Def \ref{d2.8}. Some results about the uniqueness of the
solutions of (DLR) generalizing slightly Thm \ref{t3.10} shall be reported
elsewhere (see also \cite{39,40}).

The introduced concept of the classical thermal Gibbs measure will be of
particular interest in the case of polynomial perturbations where
several solutions of the corresponding (DLR) equations may exist
\cite{18}.

Using (DLR) equation the constructive approach to the problem of
preservation of the two-sided Markov property on the circle $K_\b$ for
the limiting thermal random field $\mu_\e^\lam$ can be formulated. The
idea is to show that for $\mu_\e^\lam$ --- a.every $\Psi \in {\cal
D}' (\Rd)$ the limits:
$$
\lim_{\Lam \uparrow \Rd} E_{\mu_{\Lam,\e}^\lam} \{F | \S^0(\[t,\ s\]^C
\times \Lam^C)\}(\Psi)
$$
(where $\S\(\[t,\ s\]^C \times \Lam^C\)$ is the $\s$-algebras generated
by the random elements $\<\Phi,\ g \otimes f\>$, with $g$ supported on the
segment $\[t,\ s\]^C$ and $f$ supported in $\Lam^C$) exist and are equal
($\mu_\e^\lam $ a.e.) to the conditional expectation values:
$$
E_{\mu_\e^\lam} \left\{ F | \S \(\[t,\ s\]^C\) \right\}(\psi)
$$

Details of the proof, that indeed, for small values of $|\lam|$ this is
true, will be reported elsewhere \cite{18}.

\subsection{}
For a bounded $\Lam \subset \Rd$ the theory of bounded perturbations of
the KMS structures (see i.e. \cite{17}, Ch.\ 4 and references therein) can
be applied in the thermal representation enabling us to study the gentle
perturbations on the whole Weyl algebra. It is proven in \cite{18} that
again the nonhomogeneous thermal process $(\xi^{\lam,\Lam}_t)_{t \e
K_\b}$ determines the corresponding $W^*$-KMS structure obtained from
the corresponding GNS representation. The important problems of
constructing the perturbed (euclidean-time) Green functions on the whole
Weyl algebra ${\cal W}(\gh)$ and the questions whether the corresponding
homogenous process  $(\xi^{\lam}_t)_{t \e K_\b}$ determines them
and also whether the limiting $W^*$-KMS structure on ${\cal W}(\gh)$
forms a modular structure will be a topic of an another paper in this
series.

\subsection{}
The Abelian sector of the free Bose critical gas can be described in the
Euclidean region by certain nonergodic Gaussian generalized thermal
process. Results complementary to those contained in the section 2 for
the critical gas are obtained in \cite{18}, where also, thermodynamic
limits of the gentle perturbations on the Abelian sector have been
controlled by applications of the Fr\"ohlich-Park correlation
inequalities. The most interesting result of these investigations is
that nonergodicity of the limiting, perturbed thermal process is
preserved. Whether this is connected to the preservation of the
Bose-Einstein condensate in the interacting system remains to be
answered.

\subsection{}
More general, unbounded perturbations (i.e. of polynomial type) will be
described in an another paper of the planned series \cite{18}. Standard
tools of constructive Euclidean Quantum field theory like the high (and
the low) temperature cluster expansions are used to study the
corresponding perturbations of the free thermal structure on the Abelian
sector.

\subsection*{Acknowledgments}
Several stimulating discussions with S.\ Albeverio and Yu.\ Kondratiev
are gratefully acknowledged by R.G.\ Financial support obtained by R.G.\
from EC was very helpful.

\def\bib{\bibitem}
\def\jmp#1{{\em J.\ Math.\ Phys.\ {\bf #1}}}
\def\tmp#1{{\em Theor.\ Math.\ Phys.\ {\bf #1}}}
\def\pjm#1{{\em Pacific J.\ Math.\  {\bf #1}}}
\def\jfa#1{{\em J.\ Funct.\ Anal.\ {\bf #1}}}
\def\jsp#1{{\em J.\ Stat.\ Phys.\ {\bf #1}}}
\def\hp#1{{\em Helv.\ Phys.\ Acta {\bf #1}}}
\def\comm#1{{\em Comm.\ Math.\ Phys.\ {\bf #1}}}
\def\am#1{{\em Adv.\ Math.\ {\bf #1}}}
\def\pr#1{{\em Phys.\ Rev.\ {\bf #1}}}
\def\jetp#1{{\em Soviet Physics JETP {\bf #1}}}
\def\nc#1{{\em Nuovo Cim.\ {\bf #1}}}

\end{document}